\documentclass[a4paper,11pt]{article} 
\usepackage[margin=0.5in]{geometry} 
\usepackage{amssymb}
\usepackage{amsmath,epsfig, epsf, epic, graphicx} 
\usepackage{fancybox, lscape, subfigure} 
\usepackage{bm}
\usepackage{natbib}
\usepackage[margin=2cm]{caption}
\usepackage{setspace}
\usepackage{hyperref}

\allowdisplaybreaks

\usepackage{floatrow}
\newfloatcommand{capbtabbox}{table}[]

\title{Estimating the health effects of environmental mixtures using Bayesian semiparametric regression and sparsity inducing priors.}
\author{Joseph Antonelli, Maitreyi Mazumdar, David Bellinger, \\ David C. Christiani, Robert Wright, Brent A. Coull}

\begin{document}
\date{}
\maketitle{}


\abstract{Humans are routinely exposed to mixtures of chemical and other environmental factors, making the quantification of health effects associated with environmental mixtures a critical goal for establishing environmental policy sufficiently protective of human health. The quantification of the effects of exposure to an environmental mixture poses several statistical challenges.  It is often the case that exposure to multiple pollutants interact with each other to affect an outcome.  Further, the exposure-response relationship between an outcome and some exposures, such as some metals, can exhibit complex, nonlinear forms, since some exposures can be beneficial and detrimental at different ranges of exposure. To estimate the health effects of complex mixtures we propose a flexible Bayesian approach that allows exposures to interact with each other and have nonlinear relationships with the outcome. We induce sparsity using multivariate spike and slab priors to determine which exposures are associated with the outcome, and which exposures interact with each other. The proposed approach is interpretable, as we can use the posterior probabilities of inclusion into the model to identify pollutants that interact with each other. We utilize our approach to study the impact of exposure to metals on child neurodevelopment in Bangladesh, and find a nonlinear, interactive relationship between Arsenic and Manganese. }

\section{Introduction}

The study of the health impacts of environmental mixtures, including both chemical and non-chemical stressors, on human health is a research priority in the environmental health sciences   \citep{carlin2013unraveling,taylor2016statistical,Braun2016}. In the past, the majority of health effect studies upon which environmental regulation is based estimate the independent association between an outcome and a given exposure, either marginally or while controlling for co-exposures.     However, exposure to mixtures of chemicals is a real world scenario, and it has been argued that chemical mixture effects are possible even when each chemical exposure is below its no observable adverse effect level \citep{kortenkamp2007low}. 

Accordingly, there has recently been a renewed focus on the development of statistical methods for the analysis of the effects of chemical mixtures, or more generally environmental mixtures defined to include both chemical and non-chemical (such as diet or psychosocial factors) stressors, on health.  Recent work on this topic has recognized that investigations 
on these effects may focus on a variety of research questions of interest, including the identification of individual exposures driving this effect, identification of interactions among chemicals on health, and the estimation of the cumulative health effects of exposure to multiple agents.     Recent reviews of relevant statistical methods useful for analyzing the resulting data have highlighted that the choice of statistical method will depend on the primary objective of the study  \citep{Braun2016}, and recent work in this area
has developed both Bayesian and frequentist methods that implement forms of regularization, variable selection, dimension reduction, statistical learning, or smoothing 
in various combinations to address these questions in the presence of correlated exposures (see \cite{lazarevic2019statistical} for an excellent review). 
Notably, there is a gap in this literature in terms of methods that directly quantify the strength of evidence supporting interactions among individual chemicals on a health outcome in the data. For instance, Bayesian kernel machine regression (BKMR; \cite{Bobb2014, bobb2018statistical}) can estimate cumulative health effects of exposure to multiple agents, provide variable importance scores for each individual chemical, and estimate selected summary statistics that reflect interactions among some chemicals at given levels of others.  However, it does not yield overall variable importance scores for interactions among chemicals within the larger mixture of exposures. 

Other approaches allow for interactions between exposures, perhaps with some variable selection technique applied, but these approaches are often based on models assuming simple, linear associations between exposures or other strong parametric assumptions. It has been recently shown that in some settings, such as in studies on the impacts of exposure to metal mixtures on neurodevelopment in children, the exposure-response relationship between the multivariate exposure of interest and a health outcome can be a complex, non-linear and non-additive function \citep{henn2010early,henn2014chemical,Bobb2014}.  Therefore there is a need for statistical approaches that can handle complex interactions among a large number of exposures. Even in settings with a relatively small number of exposures, the number of potential interactions can become large, making some form of variable selection or dimension reduction necessary. In this paper we provide a Bayesian approach to this problem that allows for nonlinear effects on the outcome, can identify important exposures among a large mixture, and can identify important interactions among the elements of a mixture. 

\subsection{Review of statistical literature}

There exists a vast statistical literature on high-dimensional regression models, including  approaches to reduce the dimension of the covariate space. The LASSO  \citep{tibshirani1996regression} utilizes an L1 penalty on the absolute value of regression coefficients in a regression model, which effectively selects which variables should be included in the model by setting some coefficients equal to zero. Many approaches have built upon this same methodology \citep{zou2005regularization,zou2006adaptive}; however these all typically rely on relatively simple, parametric models. More recently, a number of papers \citep{zhao2009composite, bien2013lasso,Hao2014,lim2015learning} have adopted these techniques to identify interactions, although these approaches are only applicable to linear models with pairwise interactions. While similar ideas for nonlinear models were used by \cite{radchenko2010variable}, these analyses were restricted to pairwise interactions. Moreover, quantification of estimation uncertainty can be quite challenging in high dimensions. 

A number of approaches have been introduced that embed variable selection within nonlinear regression models. \cite{shively1999variable} utilized variable selection in Bayesian nonparametric regression based on an integrated Wiener process for each covariate in the model, and then extended these ideas in \cite{wood2002model} to non additive models that allowed for interactions between covariates. They utilized the BIC approximation to the marginal likelihood in order to calculate posterior model probabilities, which requires enumerating all potential models. Even when the number of covariates in the model is not prohibitively large, inclusion of just two-way interactions can lead to a massive number of possible models. \cite{reich2009variable} utilized Gaussian process priors for all main effects and two-way interactions within a regression model and performed variable selection in order to assess which covariates are important predictors of the outcome. They do not allow for explicit identification of higher-order interactions, as all remaining interactions are absorbed into a function of the covariates that accounts for higher-order interactions. \cite{yau2003bayesian} and \cite{scheipl2012spike} utilized variable selection in an additive regression model framework with interaction terms within a Bayesian framework, though one must specify the terms that enter the model {\em a priori} and there can be a large number of potential terms to consider. \cite{Qamar2014} used Gaussian processes to separately model subsets of the covariate space and used stochastic search variable selection to identify which covariates should enter into a given Gaussian process function, implying that they interact with each other. While this approach can be used to capture and identify complex interactions, it relies on Gaussian processes and therefore does not scale to larger data sets.

Complex models that allow for nonlinearities and interactions are becoming popular in the machine learning literature, with Gaussian process (or kernel machine) regression \citep{o1978curve}, support vector machines \citep{cristianini2000introduction}, neural networks \citep{bengio2003neural}, Bayesian additive regression trees (BART, \cite{chipman2010bart}), and random forests \citep{breiman2001random} being just a few examples. \cite{Bobb2014} used Bayesian kernel machine regression to quantify the health effects of an environmental mixture. This approach imposed sparsity within the Gaussian process to flexibly model multiple exposures. While these approaches and their many extensions can provide accurate predictions in the presence of complex relationships among exposures, they typically only provide a variable importance measure for each feature included in the model, and do not allow for formal inference on higher-order effects. An important question remains how to estimate the joint effect of a set of exposures that directly provides inference on interactions, while scaling to larger data sets without requiring a search over such a large space of potential models involving higher-order interactions. 
Further, full Markov Chain Monte Carlo (MCMC) inference in kernel machine regression does not scale with the sample size, limiting its applicability in large sample settings.  While one can potentially use approximate Bayesian methods, such as variational inference to simplify computation, these methods are only approximate and can yield biased estimators for some parameters \citep{guo2016boosting,miller2016variational}. 

\subsection{Our contribution}

In this paper we introduce a Bayesian semiparametric regression approach that addresses several of the aforementioned issues as it allows for nonlinear interactions between variables, identifies interactions between variables, and scales to larger data sets than those handled by traditional Gaussian process regression. We will introduce sparsity via spike and slab priors \citep{mitchell1988bayesian,george1993variable} within a semiparametric regression framework that reduces the dimension of the parameter space, while yielding inference on interactions among individual exposures in the mixture. Section \ref{sec:prelim} introduces the motivating data set looking at exposure to metals in Bangladesh. Section \ref{sec:model} introduces our model formulation and discusses various estimation issues. Section \ref{sec:prior} discusses some results about our prior distribution. Section \ref{sec:sims} illustrates the proposed approach through simulated data and compares it with existing approaches. Section \ref{sec:bangladesh} estimates the health effects of environmental exposures on neurodevelopment in Bangladeshi children. We will conclude with some final remarks in Section \ref{sec:discussion}.

\section{Chemical mixtures and neurodevelopment in Bangladesh}
\label{sec:prelim}
To study the health impact of chemical mixtures on neurodevelopment, we examine a study of 375 children in Bangladesh whose mothers were potentially exposed to a variety of chemical pollutants during pregnancy.  The outcome of interest is the z-score of the motor composite score (MCS), a summary measure of a child's psychomotor development derived from the Bayley Scales of Infant and Toddler Development, Third Edition (BSID-III) \citep{bayley2006bayley}. The exposure measures are log-transformed values of Arsenic (As), Manganese (Mn), and Lead (Pb) measured in umbilical cord blood, which is taken to represent prenatal exposure to these metals. As potential confounders of the neurodevelopment - exposure associations, we will also control for the following covariates: gender, age at the time of neurodevelopment assessment, mother's education, mother's IQ, an indicator for which clinic the child visited, and the HOME score, which is a proxy for socioeconomic status. Previous studies have examined these data \citep{Bobb2014,valeri2017joint} and found harmful associations between the exposures and the MCS score. Further, these studies have suggested a possible interaction between Arsenic and Manganese, though it was unclear whether this was simply random variation or a true signal. The assessment of interactions within environmental mixture data is an important problem \citep{henn2014chemical} as the identification of exposures that interact with one another can help identify groups of exposures, or exposure profiles, that have particularly strong effects on health. The ability to identify them and assign measures of uncertainty around the existence of interactions is important, and will represent an important aspect of our proposed approach.

\subsection{Preliminary analyses}

Before examining any associations with the outcome, we first look at the Pearson correlation among the log exposure values. Manganese and lead are uncorrelated, manganese and arsenic have a correlation of 0.58, and arsenic and lead have a correlation of -0.37.  Next, we evaluate marginal associations between the exposures and outcome, unadjusted for any additional covariates or co-exposures. Table \ref{tab:marginal} shows these results. We see that all three exposures have significant marginal correlations with the MCS score. We also evaluate whether a linear or nonlinear relationship between the MCS score and the exposures was favored by the data by looking at AIC values. The nonlinear AIC value is obtained by regressing the MCS score against a 3 degree of freedom spline basis representation of the exposures. We see that for Arsenic, the nonlinear model is favored, while the linear model is favored for the other two exposures. Arsenic concentrations are  positively associated with MCS score, which could possibly be due to unmeasured confounding by, say, nutrients, due to the fact that a significant exposure pathway of Arsenic exposure in this population is diet. 

\begin{table}[ht]
\centering
\begin{tabular}{lrrr}
  \hline
Exposure & Correlation with outcome & Linear AIC & Nonlinear AIC \\ 
  \hline
Arsenic & 0.46 (0.37, 0.53) & 972.67 & 966.93 \\ 
  Lead & -0.17 (-0.27, -0.07) & 1050.05 & 1053.51 \\ 
  Manganese & -0.39 (-0.47, -0.3) & 999.67 & 1001.81 \\ 
   \hline
\end{tabular}
\label{tab:marginal}
\caption{Marginal relationship between the exposures and the MCS score in the Bangladesh data. Linear AIC represents the AIC of the linear model between the MCS score and the exposures, while nonlinear AIC represents the AIC of a linear model between the MCS score and a 3 degree of freedom spline basis representation of the exposures.}
\label{tab:marginal}
\end{table}

Next, to highlight the importance of the new methodology we propose to address the presence of interactions between environmental exposures, we analyze the data using current state of the art approaches for the analysis of chemical mixtures . Specifically, we use the Bayesian Kernel Machine Regression approach (BKMR, \cite{Bobb2014}), which is now widely used within the environmental epidemiological literature  \citep{braun2017early, harley2017association}. BKMR allows for nonlinear relationships, interactions among exposures, and adjusts for additional covariates measured in the population. We use the \texttt{bkmr} package in R, generating 25,000 Markov chain Monte Carlo (MCMC) scans and the default prior specification in the package. Table \ref{tab:bkmrPIP} shows the posterior inclusion probabilities for the three metal exposures generated by this model. 

\begin{table}[ht]
\centering
\begin{tabular}{lr}
  \hline
Exposure & Posterior inclusion probability \\ 
  \hline
Pb & 0.69 \\ 
  Mn & 0.74 \\ 
  As & 0.81 \\ 
   \hline
\end{tabular}
\caption{Posterior inclusion probabilities for the three exposures analyzed in the Bangladesh data}
\label{tab:bkmrPIP}
\end{table}

\begin{figure}[ht]
\centering
	  \includegraphics[width=0.6\textwidth]{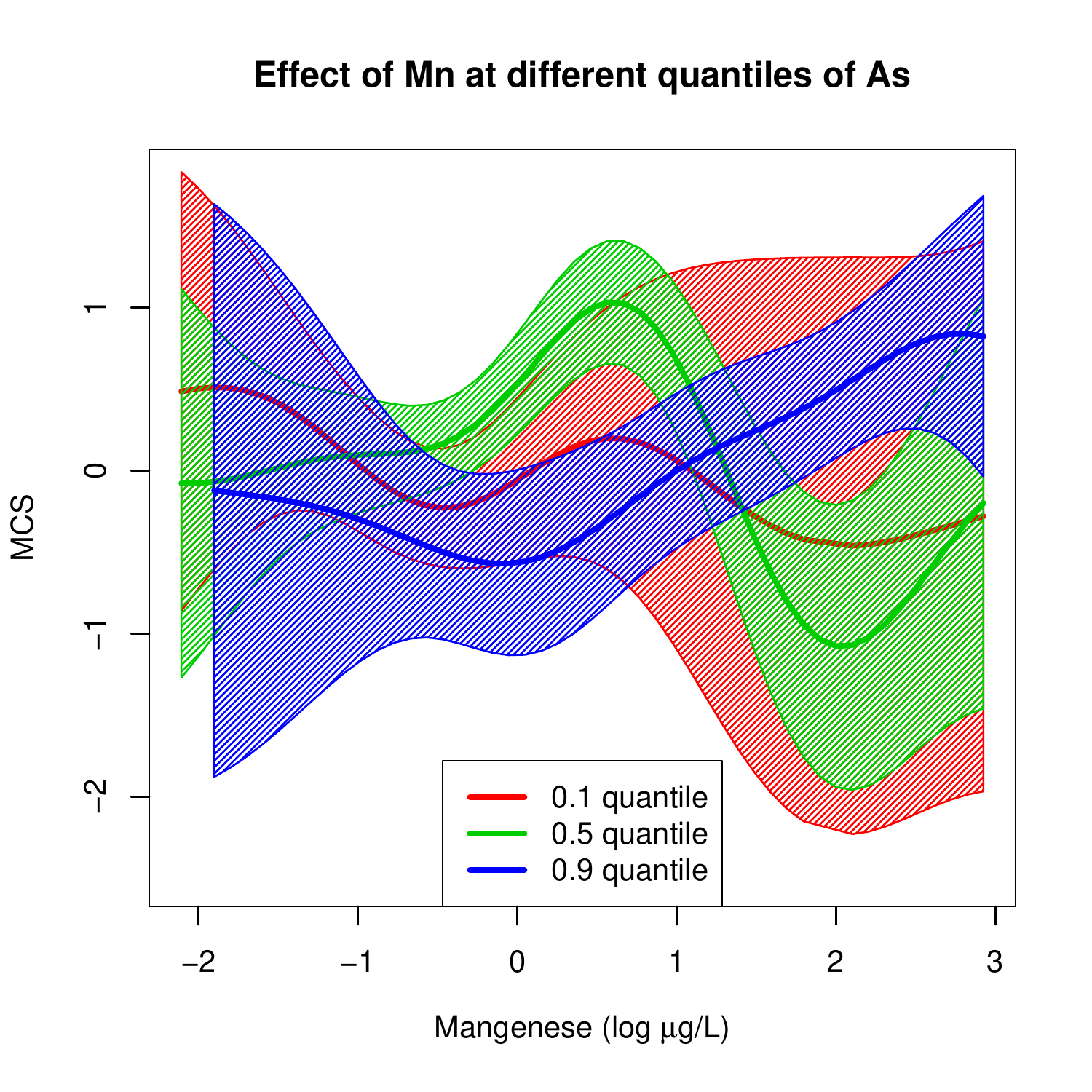} 
\caption{Exposure response curve between Manganese and MCS score at different quantiles of Arsenic.}
\label{fig:bkmrMnAs}
\end{figure}

These measures quantify the amount of uncertainty regarding the importance of each of the three exposures in predicting MCS score, but do not reflect whether or not the three exposures interact in their effects on the outcome.  One could get posterior inclusion probabilities such as these if all three exposures have distinct, important main effects, or if there is a three-way interaction between them. To begin to address this question, one can visualize the estimated exposure response curves between these exposures. We visually inspect the relationship that Manganese and Arsenic have on the MCS score. In particular, Figure \ref{fig:bkmrMnAs} shows the relationship between Manganese and the MCS score while fixing Arsenic at different quantile levels (0.1, 0.5, and 0.9). If there is an interaction between these two variables then these curves would differ across the different quantiles of Arsenic, whereas if there were no interaction, then the shape of the three curves would not differ. We see that there is some variability in the shape of the curves, particularly for the 0.5 quantile, which has an inverted U shape. This suggests the possibility of an interaction between Manganese and Arsenic; however, this can not be concluded definitively given the combination of the posterior inclusion probabilities and the exposure-response curves.  This shows the importance of new statistical methodology to identify these interactions

\section{Model formulation}
\label{sec:model}

Throughout, we assume that we observe $\boldsymbol{D_i} = (Y_i, \boldsymbol{X_i}, \boldsymbol{C_i})$ for $i,\dots,n$, where $n$ is the sample size of the observed data, $Y_i$ is the outcome, $\boldsymbol{X_i}$ is a $p$-dimensional vector of exposure variables, and $\boldsymbol{C_i}$ is an $m$-dimensional vector of covariates for subject $i$. Building on the construction of \cite{wood2006low}, we assume that we have low rank bases that represent smooth functions for each of the exposures. For instance, the main effect of $X_1$ and an interaction effect between $X_1$ and $X_2$ could be modeled as follows:

\begin{align*}
	f(X_1) = \boldsymbol{\widetilde{X}}_1 \boldsymbol{\beta}_1 \hspace{1cm} f(X_1, X_2) = \boldsymbol{\widetilde{X}}_1 \boldsymbol{\beta}_1 + \boldsymbol{\widetilde{X}}_2 \boldsymbol{\beta}_2 + \boldsymbol{\widetilde{X}}_{12} \boldsymbol{\beta}_{12},
\end{align*}

\noindent respectively, where $\boldsymbol{\widetilde{X}}_1 = [g_1(X_1), \dots, g_d(X_1)]$ and  $\boldsymbol{\widetilde{X}}_2 = [q_1(X_2), \dots, q_d(X_2)]$ represent $d-$dimensional basis function expansions of $X_1$ and $X_2$, respectively. $\boldsymbol{\widetilde{X}}_{12} = [g_1(X_1) q_1(X_2), g_1(X_1) q_2(X_2), \dots, g_d(X_1) q_d(X_2)]$ represents a $d^2-$dimensional basis expansion of the interaction between $X_1$ and $X_2$. In this work, we specify $g()$ and $q()$ to be natural spline basis functions of $X_1$ and $X_2$, respectively. For simplicity we let the degrees of freedom for each exposure all be set to $d$, though this can easily be extended to allow the degrees of freedom to vary across exposures. Importantly, the construction above enforces that main effect terms are a subset of the terms included in the interaction terms. As we include higher-order interactions, we  impose the constraint that all lower level interactions between the exposures in that interaction term are included in the basis expansion. To ease the presentation of our model, we use subscripts to denote the order of the interaction to which a coefficient applies. For instance, $\boldsymbol{\beta}_{12}$ refers to parameters associated with two-way interaction terms between $X_1$ and $X_2$ (excluding main effect terms), while $\boldsymbol{\beta}_{1}$ refers to the main effect parameters for $X_1$. This process could be trivially extended to any order interaction up to a $p-$way interaction, which is the maximum possible interaction in our model. We will use the following model formulation:

\begin{align}
	\label{eqn:Main} Y_i &\sim Normal \left(f(\boldsymbol{X}_i) + \boldsymbol{C}_i \boldsymbol{\beta}_c, \sigma^2 \right) \\
	\nonumber f(\boldsymbol{X}_i) &= \sum_{h=1}^k f^{(h)}(\boldsymbol{X}_i) \\
	\nonumber f^{(h)}(\boldsymbol{X}_i) &= \sum_{j_1=1}^p \boldsymbol{\widetilde{X}}_{i j_1} \boldsymbol{\beta}_{j_1}^{(h)} + \sum_{j_1=2}^p \sum_{j_2 < j_1} \boldsymbol{\widetilde{X}}_{i j_1 j_2} \boldsymbol{\beta}_{j_1 j_2}^{(h)} + \dots,
\end{align}

\noindent where the dots in the final line indicate that the summation extends all the way through $p-$way interactions. Our main goal is the estimation of $f(\boldsymbol{X}_i)$ and the identification of important exposures within $\boldsymbol{X}_i$. We are representing $f(\boldsymbol{X}_i)$ as the sum of $k$ different functions $f^{(h)}(\boldsymbol{X}_i)$, where $k$ is a number chosen to be sufficiently large so that all exposure effects can be captured by the model. This model is both over parameterized and not identifiable without any additional constraints. This over parameterization occurs because the model allows for all possible interactions of any order, which includes far too many parameters to reasonably estimate. It is not identifiable because each of the $k$ functions has the same functional form and the regression coefficients for a particular function $f^{(h)}(\boldsymbol{X}_i)$ are only identifiable up to a constant. To reduce the dimension of the parameter space and to identify parameters of interest, we will adopt the following prior specification for the regression coefficients: 

\begin{align}
    \nonumber P(\boldsymbol{\beta}_{S}^{(h)} \vert \boldsymbol{\zeta}) &= \left(1 - \prod_{j \in S} \zeta_{jh} \right) \delta_{\boldsymbol{0}} +  \left( \prod_{j \in S} \zeta_{jh} \right) \psi_1(\boldsymbol{\beta}_{S}^{(h)}) \\
    \nonumber & \text{where } S \text{ is some subset of } \{1,2,\dots,p\} \\
	\nonumber P(\zeta_{jh}) & = \tau_h^{\zeta_{jh}} (1 - \tau_h)^{1 - \zeta_{jh}} \boldsymbol{1}(A_h \not\subset A_m \forall m \text{ or } A_h = \{\} ) \\ 
    \nonumber & \text{where } A_h = \{j: \zeta_{jh} = 1 \}.
\end{align}


\noindent Here $ \boldsymbol{\zeta} = \left\{ \zeta_{jh} \right\} $ is a matrix of binary indicators of whether the exposure $j$ is included in the $h^{th}$ function in the model. One of our main goals is to identify important exposure effects and interactions between exposures, and these can be seen immediately from $\left\{ \zeta_{jh} \right\}$, which tells us both which exposures and which interactions are included in a particular function. We use spike and slab priors to eliminate exposures from the model, and in this paper we consider the spike to be a point mass at $\boldsymbol{0}$, while the slab, $\psi_1()$, is a multivariate normal distribution centered at $\boldsymbol{0}$ with covariance $\boldsymbol{\Sigma_{\beta}}$. We let $\boldsymbol{\Sigma_{\beta}}$ be a diagonal matrix with $\sigma^2  \sigma_{\boldsymbol{\beta}}^2$ on the diagonals. 

The prior distribution for $\zeta_{jh}$ is a Bernoulli distribution with parameter $\tau_h$. This prior is accompanied by an indicator function, $\boldsymbol{1}(A_h \not\subset A_m \forall m \text{ or } A_h = \{\} )$, which reflects the fact that each function either contains information from a unique set of exposures or is removed from the model completely. If a function only contained exposures that were all included in another function, then it would be providing redundant information and it is not needed, so we preclude this from happening by introducing this indicator function. More details on this indicator function and its implications for sampling can be found in the following section.  Full details of posterior sampling from this model can be found in the Appendix. 

\subsection{Identifiability of variable inclusion parameters}

Two scientific goals in the study of environmental mixtures are to identify which exposures are associated with the outcome and to identify interactions among exposures. To this end, our prior encodes the condition that, for any two sets of variables  $A_h$ and $A_m$ included in functions $h$ and $m$, either $A_h \not\subset A_m$ and $A_m \not\subset A_h$ or one of the two is an empty set.  An illustrative example of why this condition is necessary is a scenario in which all elements of $\boldsymbol{\zeta}$ are zero with the exception of $\zeta_{11}$ and $\zeta_{12}$. In this scenario, the first exposure has a nonzero effect in both the first and second functions of the model, i.e. $\boldsymbol{\beta}_{1}^{(1)}$ and $\boldsymbol{\beta}_{1}^{(2)}$ are both nonzero. Regardless of the values of $\boldsymbol{\beta}_{1}^{(1)}$ and $\boldsymbol{\beta}_{1}^{(2)}$, this model could be equivalently expressed with one function, rather than two. For instance, while $\boldsymbol{\beta}_{1}^{(1)} = - \boldsymbol{\beta}_{1}^{(2)}$ represents the same model as $\boldsymbol{\beta}_{1}^{(1)} = \boldsymbol{\beta}_{1}^{(2)} = \boldsymbol{0}$, inference regarding $\boldsymbol{\zeta}$ would differ across these two models. In the first scenario it would appear that the first exposure is associated with the outcome, even if in truth it is not. Our conditions avoid this scenario by ensuring that there are no functions that simply contain subsets of the exposures found in another function, and therefore $\boldsymbol{\zeta}$ will provide a meaningful and interpretable measure of the importance of each exposure and potential interaction.

Another issue affecting our model selection indicators is that low-order interactions are subsets of the models defined by higher-order interactions. For example, if the true model is such that there exists a main effect of both $X_1$ and $X_2$, but no interaction between them, we could mistakenly estimate that there is an interaction between exposure 1 and 2 simply due to the main effects. \citep{wood2006low, reich2009variable} addressed this issue in similar contexts by utilizing constraints that force the two-way interaction between $X_1$ and $X_2$ to only include information that is orthogonal to the main effects of the individual exposures. In our context, this is not so much an issue of identifiability, but rather an issue with MCMC chains getting stuck in local modes. When exploring the model space via MCMC we could accept a move that includes the two-way interaction, which is a local mode, and never be able to leave this model even if the simpler model is the true model. 

To avoid this scenario we simply need to search the model space in a particular manner. We will illustrate this in the case of a two-way interaction as follows:
$$
\boldsymbol{\zeta}_{p_1} =
\begin{pmatrix}
1&0&0&0\\
0&0&0&0\\
0&0&0&0\\
\end{pmatrix}
\hspace{0.5cm}
\boldsymbol{\zeta}_{p_2} = 
\begin{pmatrix}
1&0&0&0\\
1&0&0&0\\
0&0&0&0\\
\end{pmatrix}
\hspace{0.5cm}
\boldsymbol{\zeta}_{p_3} = 
\begin{pmatrix}
1&0&0&0\\
0&1&0&0\\
0&0&0&0\\
\end{pmatrix}
$$
Imagine that at a given iteration of an MCMC, $\boldsymbol{\zeta} = \boldsymbol{\zeta}_{p_1}$ and we look to update $\zeta_{21}$, which indicates whether exposure 2 enters into function 1. If we do this naively, we simply could compare models $\boldsymbol{\zeta}_{p_1}$ and $\boldsymbol{\zeta}_{p_2}$, which would result in a choice of  $\boldsymbol{\zeta}_{p_2}$ even if there were only main effects of $X_1$ and $X_2$ and no interaction between them. If, however, we update $\zeta_{22}$ at the same time, we also evaluate model $\boldsymbol{\zeta}_{p_3}$. If the true relationship is a main effect of the two exposures, then model $\boldsymbol{\zeta}_{p_3}$ will be favored by the data and we will estimate $\boldsymbol{\zeta}$ accurately.  If the truth is a two-way interaction between $X_1$ and $X_2$ then $\boldsymbol{\zeta}_{p_2}$ will be favored by the data. This setting generalizes to higher-order interactions and illustrates how we can avoid misleading conclusions about interactions. Whenever we propose a move to a higher-order interaction, we must also evaluate all lower-order models that could lead to incorrectly choosing the higher-order model. While proposing moves in this manner keeps the MCMC from getting stuck at incorrect models, it needs to be carefully constructed to maintain a valid sampler. In the Appendix, we prove that this strategy can be tailored in such a way that the sampler is reversible and therefore maintains the correct target distribution.

\subsection{Selection of spline degrees of freedom}

An important tuning parameter in the proposed model is the number of degrees of freedom, $d$, used to flexibly model the effects of the $p$ exposures. Throughout we will make the simplifying assumption that the degrees of freedom is the same for all exposures, though this can be relaxed to allow for differing amounts of flexibility across the marginal functions. We propose the use of the WAIC model selection criterion \citep{watanabe2010asymptotic, gelman2014bayesian} to select $d$. If we let $s=1,\dots,S$ index samples from the posterior distribution of the model parameters and let $\boldsymbol{\theta}$ represent all parameters in the model, the WAIC is defined as
\begin{align*}
WAIC = -2 \left(\sum_{i=1}^n \log \left( \frac{1}{S} \sum_{s=1}^S p(Y_i \vert \boldsymbol{\theta}^s) \right) - \sum_{i=1}^n \text{var}(\log p(Y_i \vert \boldsymbol{\theta})) \right).
\end{align*}

\noindent The WAIC approximates leave one out cross validation, and therefore provides 
an assessment of model fit based on out of sample prediction performance. We  first draw samples from the posterior distributions of the model parameters under a range of values for $d$, and then select the model that minimizes WAIC. While this is not a fully Bayesian approach to choosing $d$, we will see in Section \ref{sec:sims} that this approach still leads to credible intervals with good frequentist properties.

\subsection{Prior choice}
\label{sec:priorelicit}
There are a number of hyperparameters that need to be chosen to complete the model specification. First, one must choose an appropriate value for $\boldsymbol{\Sigma_{\beta}}$, which controls the amount of variability and shrinkage of $\boldsymbol{\beta}_{S}^{(h)}$ for those coefficients originating from the slab component of the prior. We will make the simplifying assumption that $\boldsymbol{\Sigma_{\beta}}$ is a diagonal matrix with $\sigma^2  \sigma_{\boldsymbol{\beta}}^2$ on the diagonals. It is well known that variable selection is sensitive to the choice of this parameter \citep{reich2009variable}, so it is crucially important to select an appropriate value. We propose the use of an empirical Bayes strategy where we estimate $\sigma_{\boldsymbol{\beta}}^2$ and then obtain posterior samples conditional on this value. A description of the algorithm for finding the empirical Bayes estimate is available in the Appendix. One could alternatively put a hyper prior on $\sigma_{\boldsymbol{\beta}}^2$ for a fully Bayesian analysis, though we have found that the empirical Bayes strategy works better in practice, so we restrict attention to that case. 

We also must select a value of $\tau_h$, the prior probability of an exposure being included in function $h$. We will assume that $\tau_h \sim \text{Beta}(M,\gamma$), and set $M$ and $\gamma$ to fixed values. One can choose values to induce a desired level of sparsity in the model, however in practice, the appropriate degree of sparsity is typically not known and therefore, in order to accommodate higher-dimensional data,  we choose values that scale with the number of exposures. We will set $\gamma = p$, which allows the prior amount of sparsity to increase as the number of exposures increases. This strategy closely resembles other existing approaches in high-dimensional Bayesian models, such as those proposed by \cite{Zhou2014} and \cite{rovckova2016spike}. We will discuss the implications of scaling the prior with $p$ in Section \ref{sec:prior}. 

The only remaining prior distributions to specify are for $\boldsymbol{\beta}_C$ and $\sigma^2$. For both parameters we specify diffuse, conjugate priors. In the case of $\boldsymbol{\beta}_C$ this means independent normal prior distributions with very large variances. For the residual variance $\sigma^2$ we specify an inverse gamma distribution with both parameters set to $0.001$.

\subsection{Lower bound on slab variance}

One issue with the aforementioned empirical Bayes estimation of $\sigma_{\boldsymbol{\beta}}^2$ is that it will be estimated to be very small if there is little to no association between the exposures and the outcome. This result can lead to problems in the estimation of $\boldsymbol{\zeta}$. If the prior variance for the slab is very small, then it is almost indistinguishable from the spike, since in this case it approximates a point mass at zero. In this situation, updating $\zeta_{jh}$ becomes essentially a coinflip for all $j$ and $h$, which will lead to high posterior inclusion probabilities even when there is no true association. To avoid this, we estimate a lower bound, above which we believe that any high posterior inclusion probabilities reflect true signals in the data. 

To estimate this lower bound, we first permute the rows of $Y$, thereby breaking any associations between the exposures and the outcome. We run our MCMC algorithm for a small number of iterations on the permuted data and keep track of the probability that $\zeta_{jh} = 1$ when $\tau_h = 0.5$ for all $h$. We do this for a decreasing grid of potential values for $\sigma_{\boldsymbol{\beta}}^2$ and we take as our lower bound for $\sigma_{\boldsymbol{\beta}}^2$ the smallest value of $\sigma_{\boldsymbol{\beta}}^2$ such that a pre-chosen threshold is met. Examples of such criteria include the smallest value of $\sigma_{\boldsymbol{\beta}}^2$ such that the probability of including a main effect for any of the exposures is less than 0.25 or the smallest value such that the probability of any two-way interaction is less than 0.05. Alternatively, because one can calculate the average false discovery rate for each value of $\sigma_{\boldsymbol{\beta}}^2$, 
one can use this strategy to implement Bayesian counterparts to traditional multiple testing adjustments that control frequentist error rates, such as the false discovery rate. It is important to note that this false discovery rate control is conditional on a fixed value of $\tau_h$, which we choose to be 0.5. In practice, we let $\tau_h$ be random and estimated from the data. Nonetheless, this procedure should provide a reasonable lower bound for $\sigma_{\boldsymbol{\beta}}^2$. If strict control of false discovery rate is required then one could specify a truncated prior for $\tau_h$ such that it is never greater than the value of $\tau_h$ used for the calculation of the lower bound (i.e., 0.5). It is important to note that this procedure is very fast computationally, generally taking only a small fraction of the computation time as the full MCMC for one data set. 

\begin{figure}[ht]
\centering
	  \includegraphics[width=0.5\linewidth]{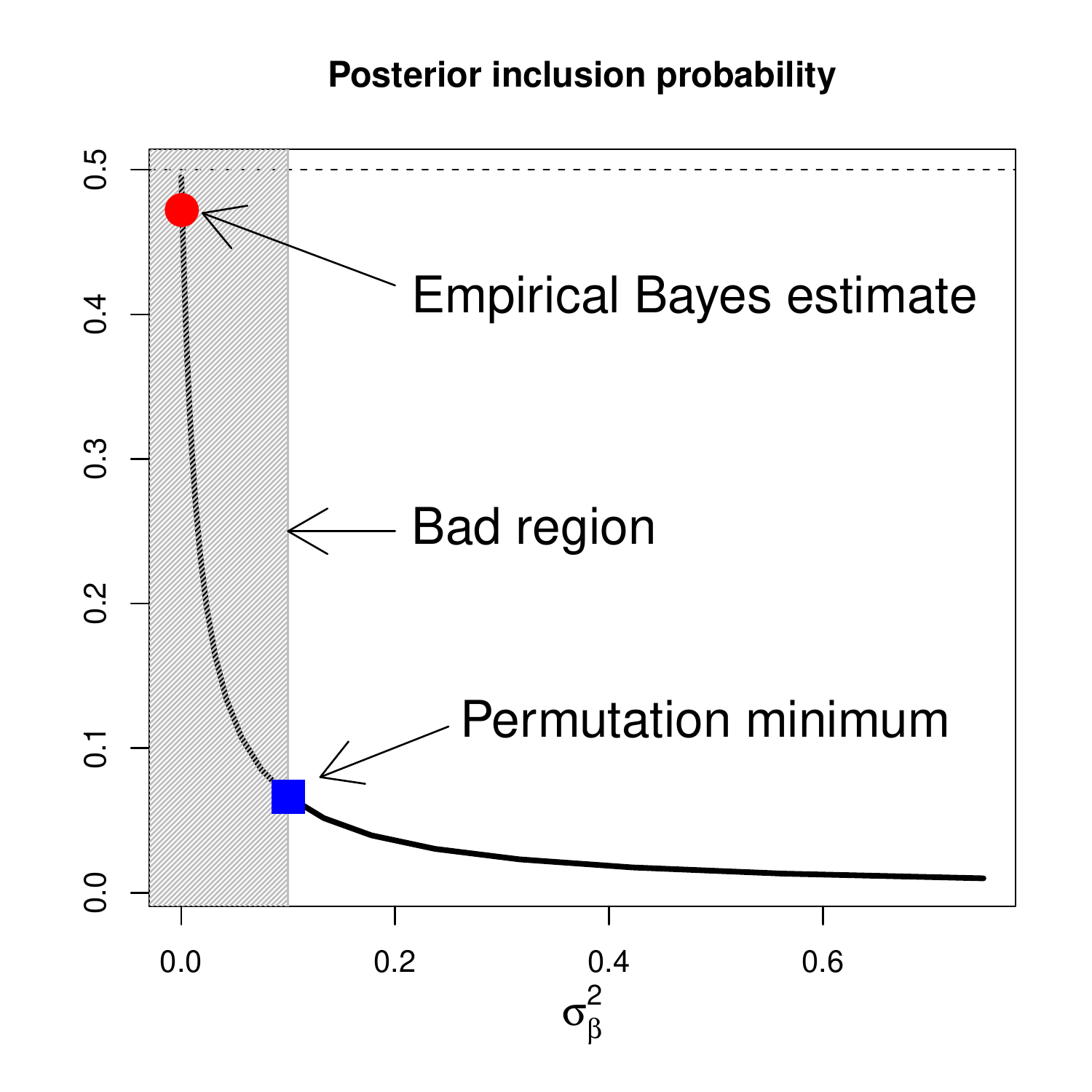}
\caption{Illustration of empirical Bayes estimate and lower bound estimate of $\sigma_{\boldsymbol{\beta}}^2$}
\label{fig:NullProb}
\end{figure}

To illustrate how this approach works for a given analysis, we simulated data with no associations between the exposure and outcome. In this case, the empirical Bayes estimate of $\sigma_{\boldsymbol{\beta}}^2$ will be very small as it is roughly estimated to be the average squared value of the nonzero coefficients in $\boldsymbol{\beta}$. Figure \ref{fig:NullProb} shows the posterior inclusion probabilities averaged over all exposures in this example as a function of $\sigma_{\boldsymbol{\beta}}^2$. The gray region represents the area where false positives start to arise because $\sigma_{\boldsymbol{\beta}}^2$ is too small, and we see that the empirical Bayes estimate is contained in this region. If we proceed with the empirical Bayes estimate, then we would conclude that all of the exposures had posterior inclusion probabilities near 0.5, incorrectly indicating they are important for predicting the outcome. Our permutation strategy to identifying a lower bound is able to find a value of $\sigma_{\boldsymbol{\beta}}^2$ that avoids high posterior inclusion probabilities for exposures with null associations with the outcome. In practice, we estimate both the lower bound and empirical Bayes estimate and proceed with the maximum of these two values, which in this simulated case is the lower bound. 

\section{Properties of Prior Distributions}
\label{sec:prior}

A primary goal in many analyses of the health effects of an environmental mixture is to identify interactions among exposures on risk. Therefore, it is important to understand the extent to which the proposed prior specification induces shrinkage on the interaction terms. Both $\tau_h$ and $\boldsymbol{\Sigma_{\beta}}$ dictate the amount of shrinkage.  We utilize empirical Bayes to select $\boldsymbol{\Sigma_{\beta}}$. Therefore, here we focus on the role of $\tau_h$, which directly controls the probability that a $j^{\text{th}}$ order interaction is nonzero in the model. The prior for this parameter can be controlled through selection of $M$ and $\gamma$. We examine the probability that a $j^{\text{th}}$ order interaction is zero by deriving the probability that the first $j$ covariates are included in an interaction term, as this probability will be the same for any other interaction of the same order. \\

\textit{Lemma 3.1:} Assuming Model \ref{eqn:Main}, the prior probability that there does not exist a function, $k'$ in the model such that $\zeta_{1k'} = \zeta_{2k'} = \dots = \zeta_{jk'} = 1$ and $\zeta_{j+1k'} = \dots = \zeta_{pk'} = 0$ is
\begin{align}
	\left( 1 - \frac{\Gamma(j + M)\Gamma(p + \gamma - j)}{\Gamma(p + \gamma + M)}\right)^k.
    \label{eq:lemma31}
\end{align}

The Appendix contains a proof of this result, which is useful because it expresses the probability of an interaction of any order being zero as a function of $M$ and $\gamma$. This result can guide the user in specifying values of $M$ and $\gamma$ that best represents prior beliefs. It has been noted in the literature that a desirable feature of the prior is to place more shrinkage on higher-order interactions \citep{gelman2008weakly,Zhou2014}. Therefore we use (\ref{eq:lemma31}) to derive conditions that accomplish this type of shrinkage. Specifically, Theorem 3.2 provides a default choice of $\gamma$ that specifies the amount of shrinkage on an interaction term to be an increasing function of the order of that term. \\

\textit{Theorem 3.2:} Assuming Model \ref{eqn:Main}, such that $\tau_h \sim \text{Beta}(M, \gamma)$, then the probability in Lemma 3.1 increases as a function of $j$ if $\gamma \geq p + M - 1$ \\

The Appendix contains a proof of this result, which shows that higher-order interactions are more aggressively forced out of the model under either the standard $\text{Beta}(1, p)$ prior, which is often used in high-dimensional Bayesian models \citep{Zhou2014,rovckova2016spike},  or if $\gamma \geq p + M - 1$, which shrinks higher-order interactions even more aggressively than this beta prior. In practice, a common prior choice for $\tau_h$ is $\text{Beta}(M, p)$, with $M$ being some constant that does not change with the number of exposures. Nonetheless, this result shows that one can achieve stronger penalization of higher-order interaction terms as long as we scale the prior with $p$.

\section{Simulation study}
\label{sec:sims}
Here we present results from a simulation study designed to assess the operating characteristics of the proposed approach in a variety of settings. In all cases we take the maximum of the empirical Bayes and lower bound estimates of $\sigma_{\boldsymbol{\beta}}^2$. For each simulated data set, we fit the model for values of $d \in \left\{ 1,2,3,4,5,6 \right\}$, and run a Gibbs sampler for 10,000 iterations with two chains, keeping every 8th sample and discarding the first 2,000 as burn-in. We present results for each value of $d$, the value that is selected based on the WAIC, the spike and slab GAM (ssGAM) approach of \cite{scheipl2012spike}, and for the Bayesian kernel machine regression (BKMR) approach of \cite{Bobb2014}. We assess the performance of the various models by calculating the mean squared error (MSE) and 95\% credible interval coverages of the true relationship between the exposures and the outcome, $f(X_1,...,X_p)$, at all observed values of the exposure.   We  also assess the ability of the proposed model to identify nonzero exposure effects and interaction terms. 

\subsection{Polynomial interactions}
First, we generate data with a sample size of $n = 200$, and generate $m=10$ covariates, $\boldsymbol{C}$, from independent standard normal distributions. We generate $p=10$ exposures, $\boldsymbol{X}$, from a multivariate normal distribution with mean $\boldsymbol{\mu}_X$. We will let the mean of $\boldsymbol{X}$ be a linear function of the covariates, i.e. $\boldsymbol{\mu}_X = \boldsymbol{C \alpha}$. We let all pairwise correlations between the exposures be 0.6, which is the average pairwise correlation among the exposures in the NHANES data analyzed in Appendix A. The marginal variances are set to 1 for all exposures. We set the residual variance to $\sigma^2 = 1$. For the outcome, we use a polynomial model that contains both a linear and a nonlinear interaction term as follows:
\begin{align*}
	Y = 0.7 X_2 X_3 + 0.6 X_4^2 X_5 + \boldsymbol{C \beta}_C + \epsilon
\end{align*}

For exact values of $\boldsymbol{\alpha}$ and $\boldsymbol{\beta}_C$ see Appendix B. Figure \ref{fig:simres1} presents the results of the simulation study. The top left panel of Figure \ref{fig:simres1} shows the matrix of posterior probabilities that a pair of exposures interact with each other. Specifically, cell $(i, j)$ of this heatmap shows the posterior probability that exposures $i$ and $j$ interact with each other in the model averaged over all simulations. Results show that the model correctly identifies the pairs $(X_2, X_3)$ and $(X_4, X_5)$ as interacting with each other. The top right panel shows the average 95\% credible interval coverage of $f(X_i)$ for each approach considered. Our model based on WAIC achieves nearly the nominal coverage indicating that we are accounting for all sources of uncertainty. The bottom left panel highlights the ability of our procedure to estimate the marginal effect of $X_4$, which has a quadratic effect on the outcome, while holding constant the values of other exposures. The grey lines, which represent individual simulation estimates, reflect the true relationship, denoted by the red line. Finally, the bottom right panel of Figure \ref{fig:simres1} presents the mean squared error of the global effect estimates of $f(X_i)$. Results show that well chosen values of $d$, including the automated choice using WAIC, lead to the best performance among the methods considered.

\begin{figure}[ht]
\centering
	  \includegraphics[height=0.255\textheight]{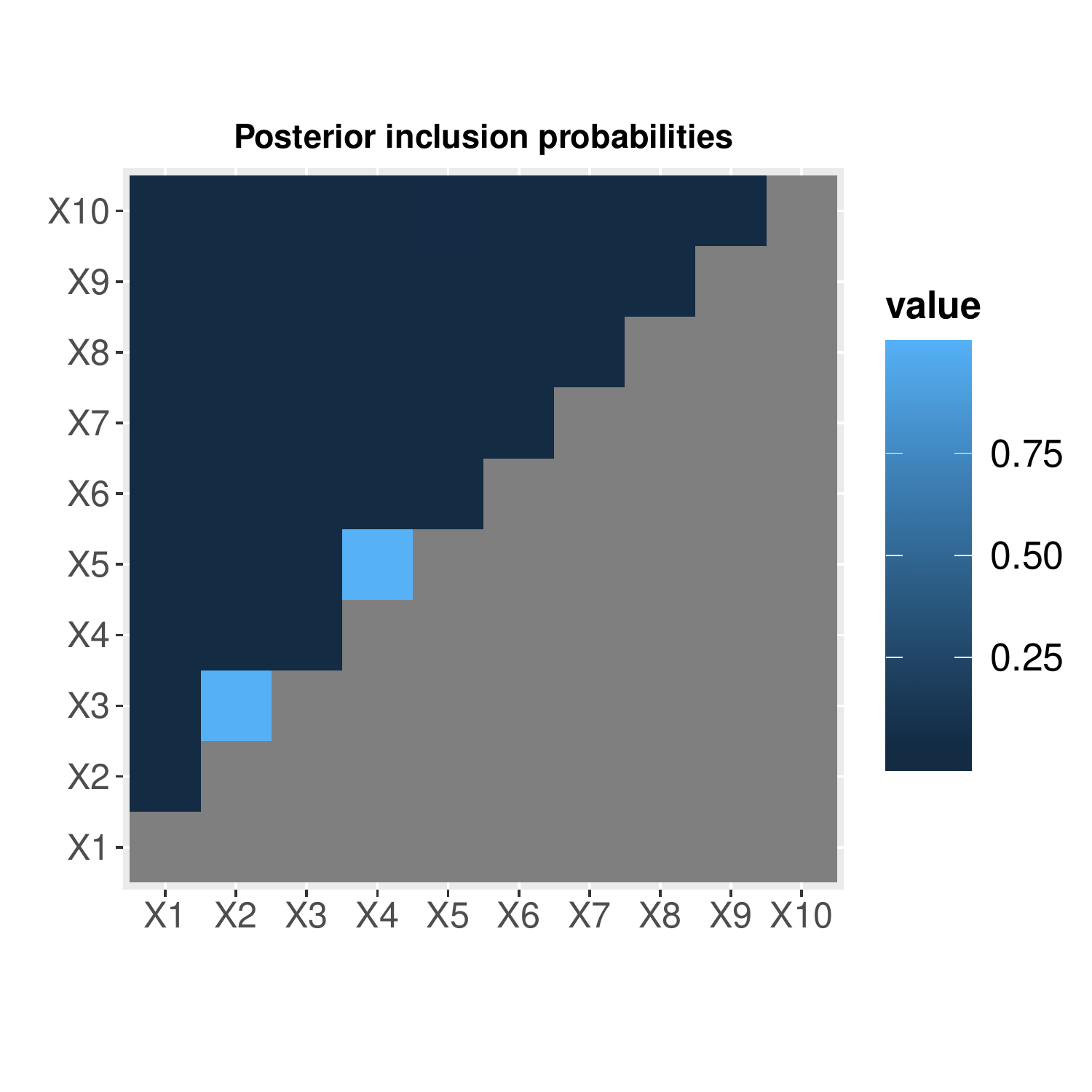} 
      \includegraphics[height=0.245\textheight]{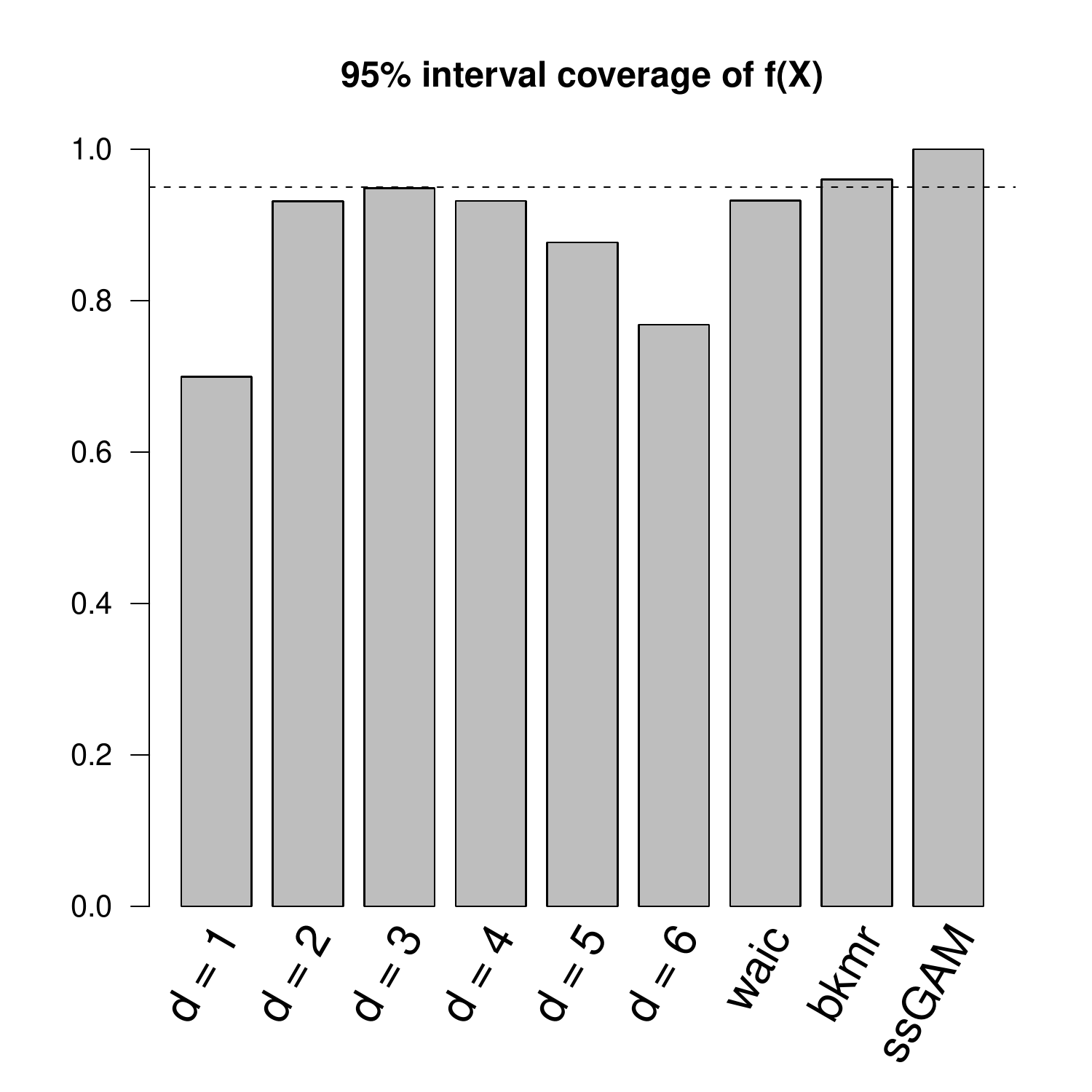} \\
      \includegraphics[height=0.25\textheight]{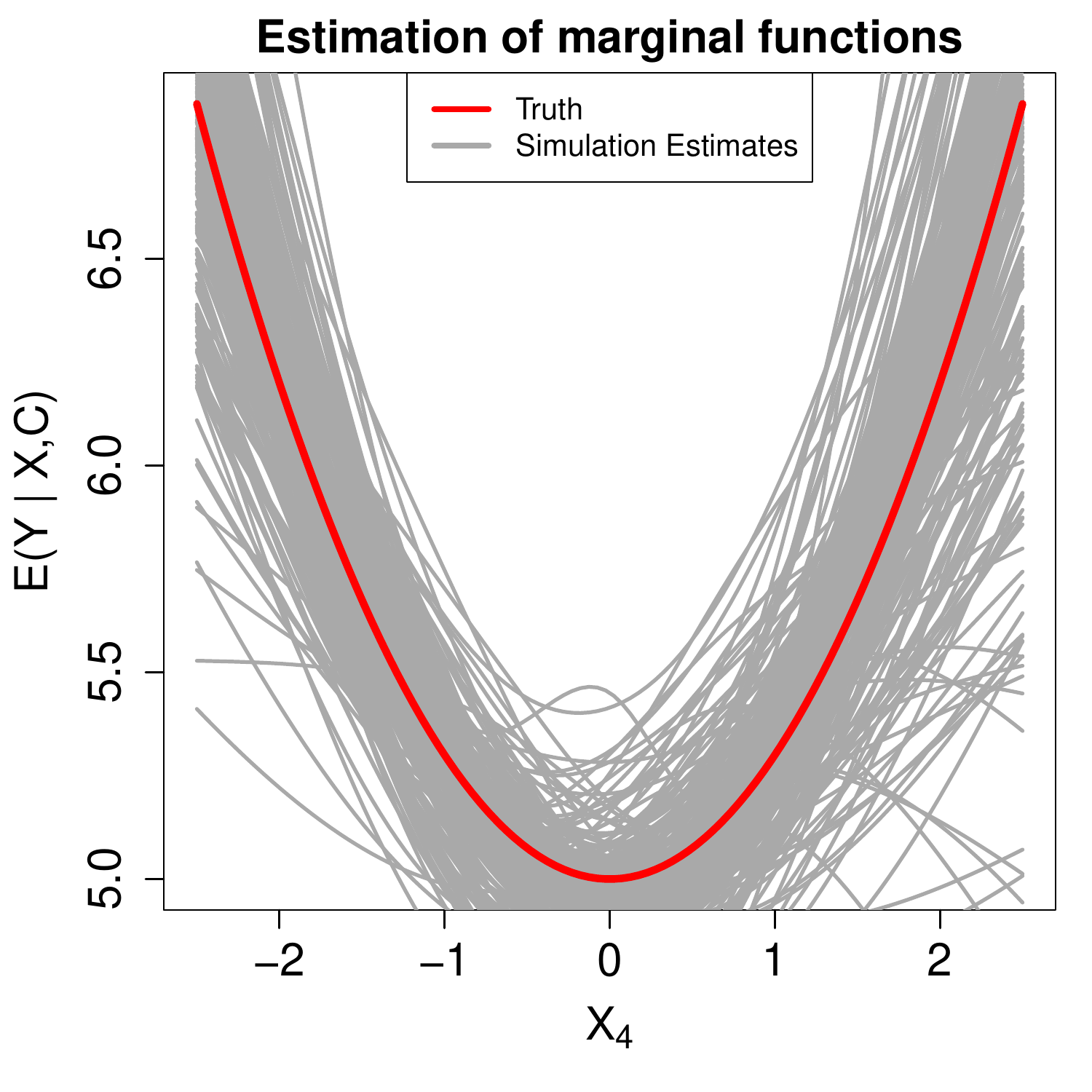}
      \includegraphics[height=0.247\textheight]{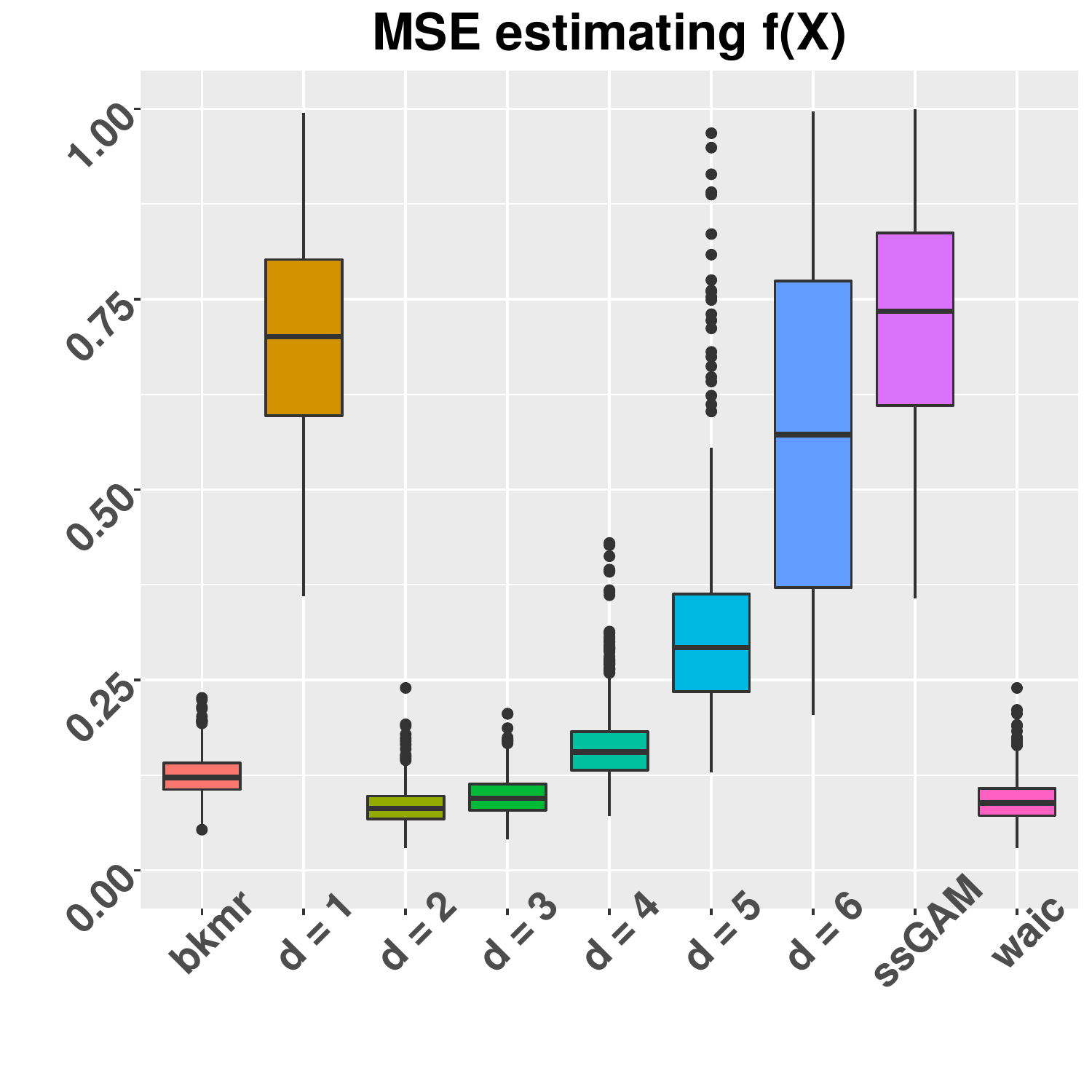}
\caption{Results of the quadratic simulation. The top left panel shows the posterior probabilities of exposures interacting with each other for the model selected by the WAIC. The top right panel shows the 95\% credible interval coverages of $f(X_i)$. The bottom left panel shows the estimated marginal functions of $X_4$ on Y for the model selected by the WAIC, where the red line represents the truth and the grey lines represent individual estimates from separate simulations. The bottom right panel shows the MSE in estimating $f(\boldsymbol{X})$ for all models considered.}
\label{fig:simres1}
\end{figure}

\subsection{High-dimensional data with three-way interaction}
\label{sec:simHD}

Now we present a scenario with $n = 10,000$ and $p=100$ to highlight that our approach can handle large sample sizes and large exposure dimensions. 
We will generate the exposures from independent uniform random variables. We generate one additional covariate, $C$, from a standard normal distribution, and we set the residual variance to $\sigma^2 = 1$. The outcome model is closely related to one that was explored in \cite{Qamar2014} and corresponds to a model that includes a three-way, nonlinear interaction as follows:

\begin{align*}
	Y = 2.5 \sin(\pi X_1 X_2) + 1.5 \cos(\pi (X_3 X_4 + X_5)) + 2(X_6 - 0.5) + 2.5X_7 + \epsilon
\end{align*}

\begin{figure}[ht]
\centering
	  \includegraphics[height=0.24\textheight]{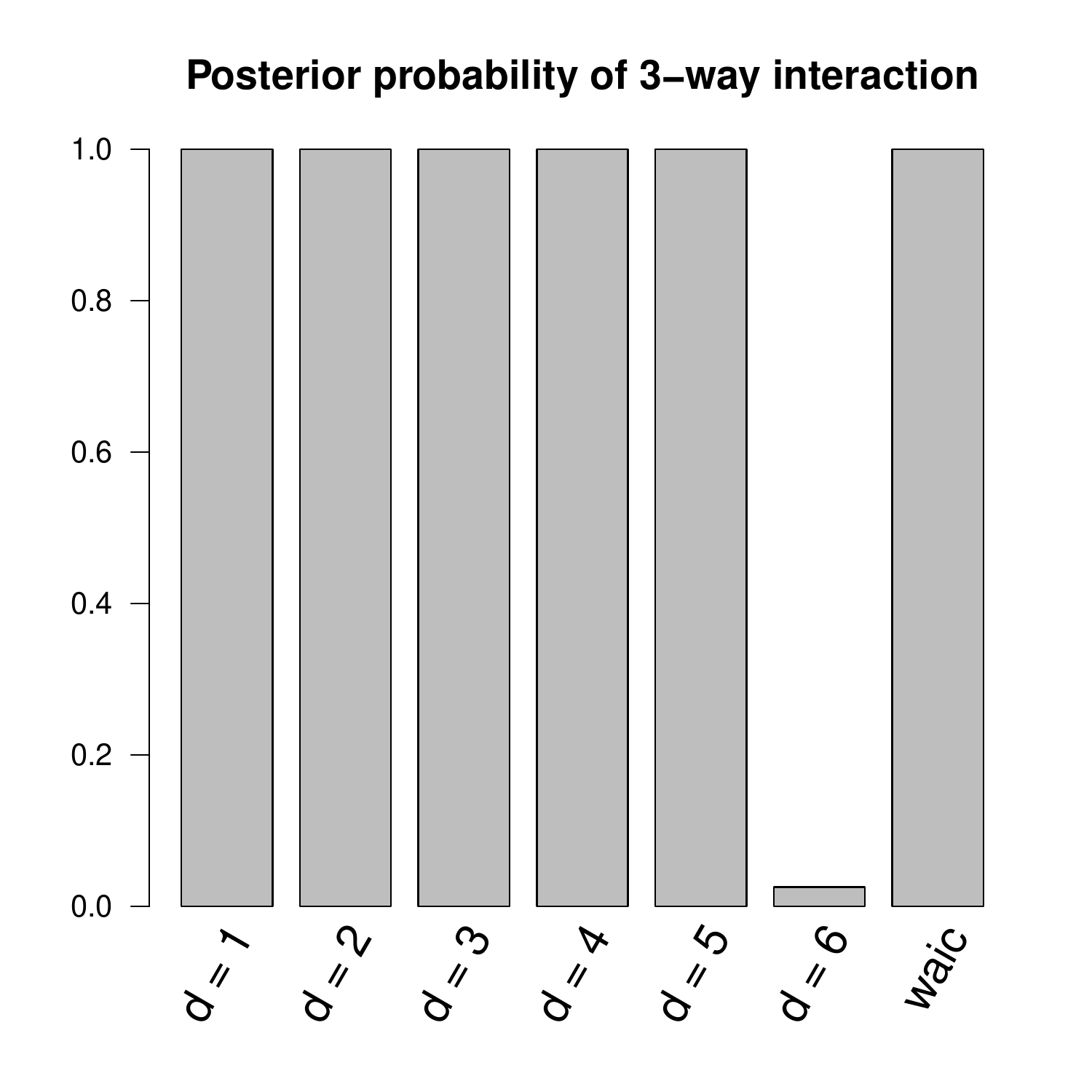} 
      \includegraphics[height=0.24\textheight]{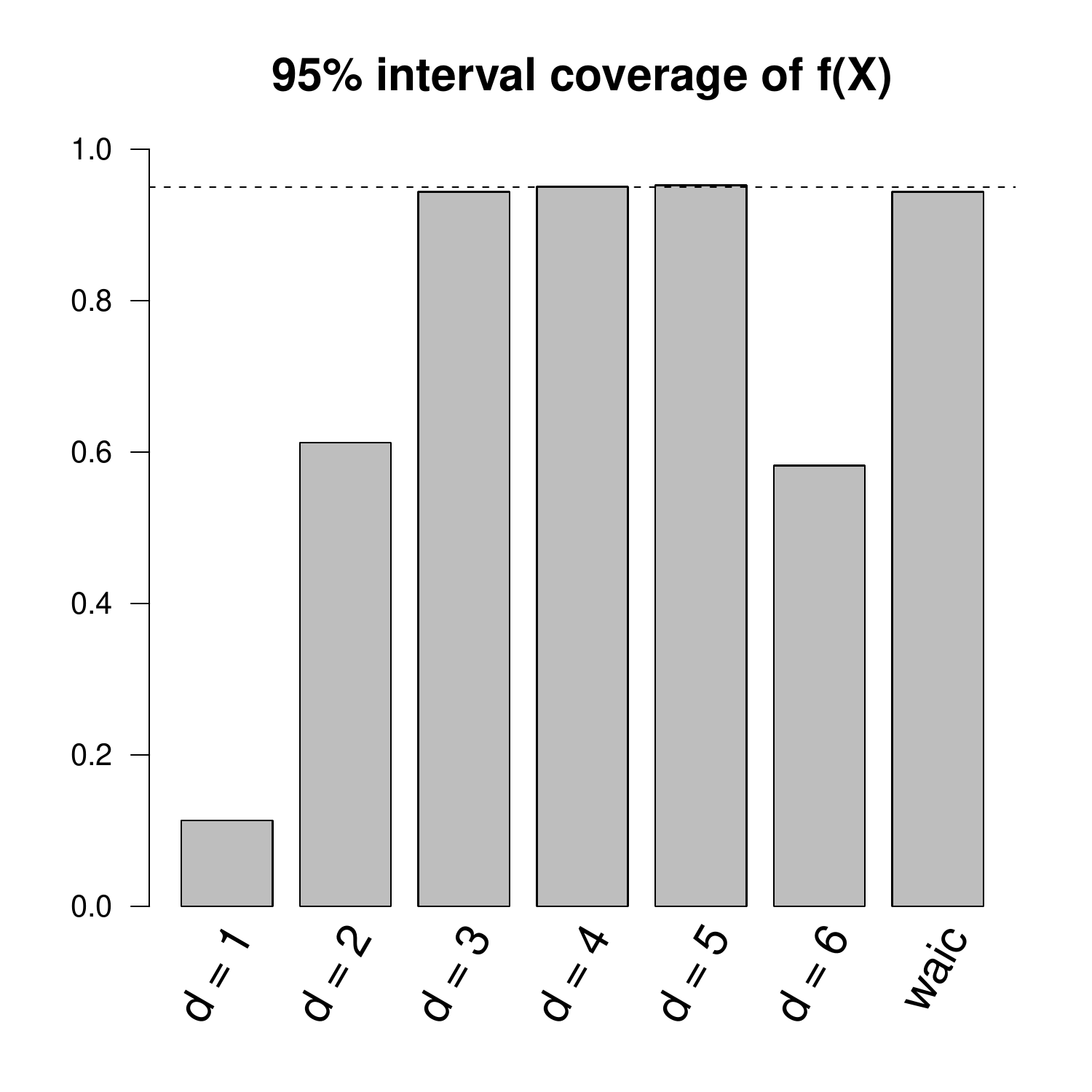} \\
      \includegraphics[height=0.25\textheight]{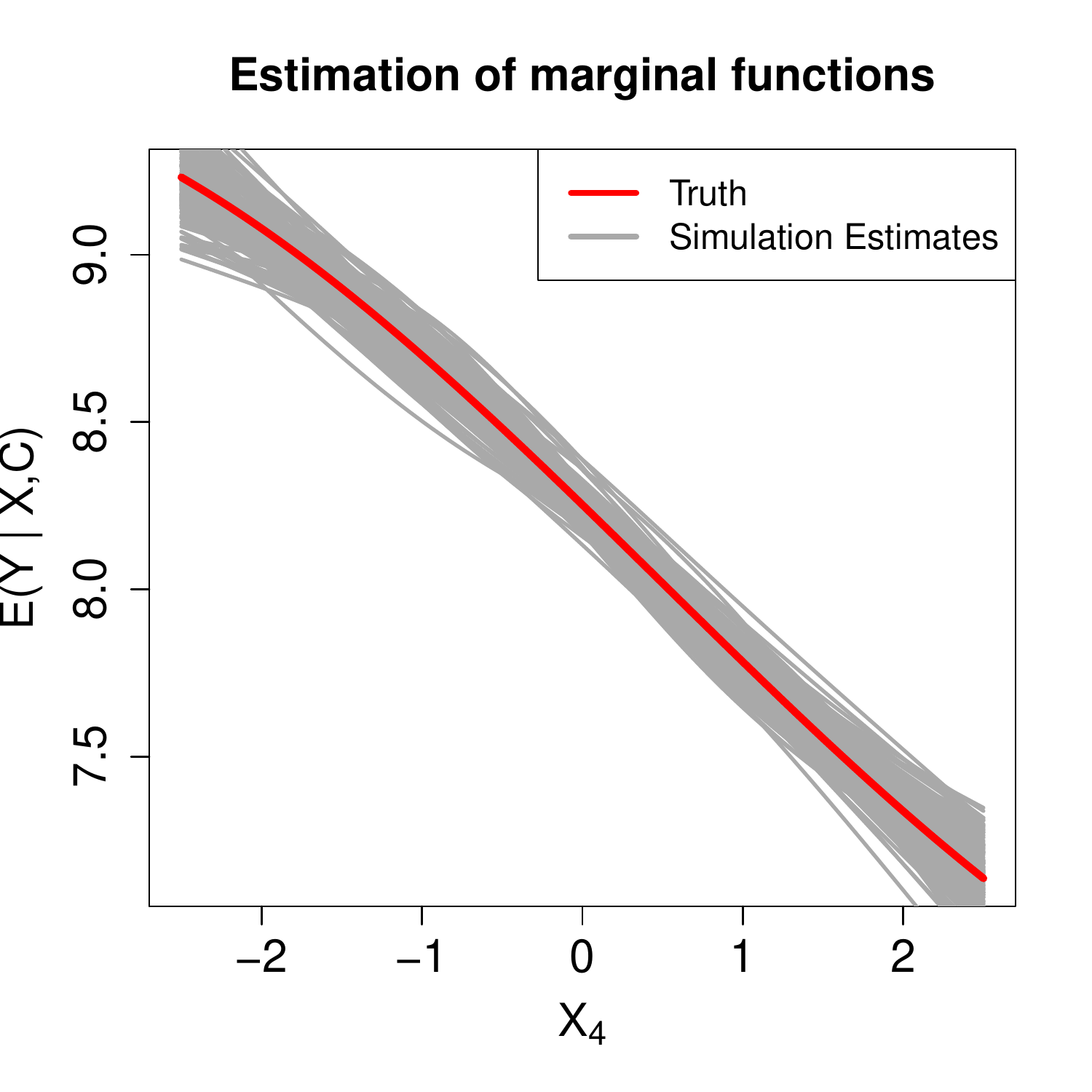}
      \includegraphics[height=0.247\textheight]{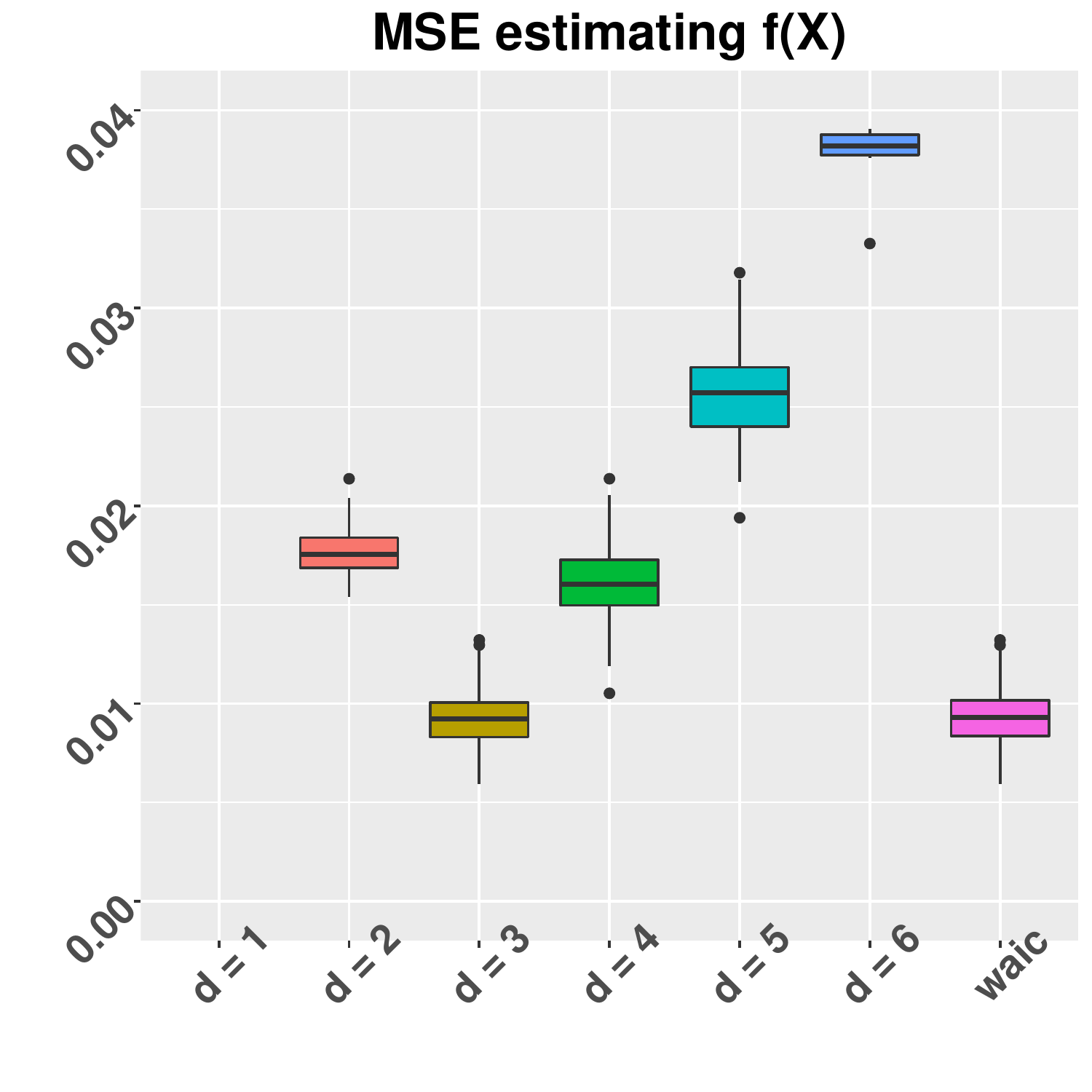}
\caption{Results of the high-dimensional simulation. The top left panel shows the posterior probabilities of the 3-way interaction between $X_3, X_4$ and $X_5$ for all models. The top right panel shows the 95\% credible interval coverages of $f(X_i)$. The bottom left panel shows the estimated marginal functions of $X_4$ on Y for the model selected by the WAIC, where the red line represents the truth and the grey lines represent individual estimates from separate simulations. The bottom right panel shows the MSE in estimating $f(\boldsymbol{X})$ for all models considered. The MSE for $d=1$ is too large to be seen on the plot.}
\label{fig:simres2}
\end{figure}

\noindent  It is clear that the relationship between the exposures is nonlinear and non-additive. We do not show results for BKMR due to the sample size limitation from using Gaussian processes. Further, there are a large number of exposures considered here, as there are 4,950 possible two-way interactions and 161,700 possible three-way interactions. For this reason, we do not include ssGAM in this simulation as it requires specifying the form of the linear predictor, and there are too many terms to consider. Our approach considers interactions of any order, though it does not require pre-specification of which terms to consider, and aims to find the correct model among all possible interactions. Figure \ref{fig:simres2} shows results from this scenario. The top left panel shows the posterior probability of a three-way interaction between $X_3, X_4$ and $X_5$. This probability is 1 for all models except for $d=6$, which does not capture this interaction. The top right panel shows the 95\% credible interval coverages of $f(X_i)$. We see that the coverage is low for $d=1$ and $d=2$, which are not sufficiently flexible to capture the true function, while the coverage is low for $d=6$ since it does not capture the three-way interaction. Importantly, the model chosen by WAIC achieves the nominal coverage level. The bottom left panel of Figure \ref{fig:simres2} shows the estimated marginal effect of $X_4$ on the outcome, fixing the other exposures. Results suggest that in almost all cases the method accurately estimates the true effect, which is close to linear in the range of the data we are looking at. The bottom right panel of Figure \ref{fig:simres2} shows that the MSE of the model chosen by WAIC is competitive, or better, than any individual choice of $d$, suggesting that WAIC is an effective method for choosing the degrees of freedom for the spline terms.

\section{Analysis of Bangladesh data}
\label{sec:bangladesh}

Here we analyze the data on metal mixtures and neurodevelopment introduced in Section \ref{sec:prelim}. We applied our proposed approach to identify the complex relationship between the three metal exposures and neurodevelopment. We ran the Markov chain for 40,000 iterations with $M=3, \gamma = p, a_0=0.001$, $b_0=0.001$, and a non-informative, normal prior on $\boldsymbol{\beta_c}$. We ran the model for values of $d \in \left\{ 1,2,3,4,5 \right\}$. Convergence of the MCMC was confirmed using trace plots and PSR values, and these results can be seen in the Appendix.

\subsection{The importance of the degrees of freedom}

Here we evaluate the impact that the degrees of freedom $d$ had on the analysis. Table \ref{tab:bangladeshWAIC} shows the WAIC values for each of the values of $d$ considered, the corresponding main effect posterior inclusion probabilities for each exposure, as well as the posterior inclusion probability for the two-way interaction between Manganese and Arsenic. The WAIC values strongly indicate that linear models are not appropriate for this data, as higher degrees of freedom lead to lower WAIC values and therefore better model fits. One of the most telling takeaways from Table \ref{tab:bangladeshWAIC} is the correspondence between WAIC values and the posterior inclusion probabilities for the two-way interaction between Manganese and Arsenic. Models with higher degrees of freedom lead to lower WAIC values and higher posterior inclusion probabilities for the two-way interaction. It appears that the lower values of $d$ are not adequate for capturing the interaction between Manganese and Arsenic, and this leads to poorer fits to the data. The marginal inclusion probabilities for each exposure are more stable across $d$, however, these go up slightly from $d=1$ to $d=5$ as well. Overall, the WAIC indicates that $d=5$ provides the best fit to the data and we will therefore present results from this model from this point onwards. 

\begin{table}[ht]
\centering
\begin{tabular}{rrrrrr}
  \hline
d & WAIC & Pb & Mn & As & Interaction between Mn and As \\ 
  \hline
  1 & 922.79 & 0.83 & 0.79 & 0.86 & 0.21 \\ 
    2 & 922.16 & 0.76 & 0.82 & 0.97 & 0.43 \\ 
    3 & 910.77 & 0.78 & 0.99 & 1.00 & 0.82 \\ 
    4 & 902.96 & 0.94 & 0.98 & 1.00 & 0.96 \\ 
    5 & 901.39 & 0.92 & 1.00 & 1.00 & 0.94 \\ 
   \hline
\end{tabular}
\caption{WAIC and posterior inclusion probabilities for differing degrees of freedom.}
\label{tab:bangladeshWAIC}
\end{table}

\subsection{Posterior inclusion probabilities}

\begin{figure}[ht]
\centering
\includegraphics[height=2.5in]{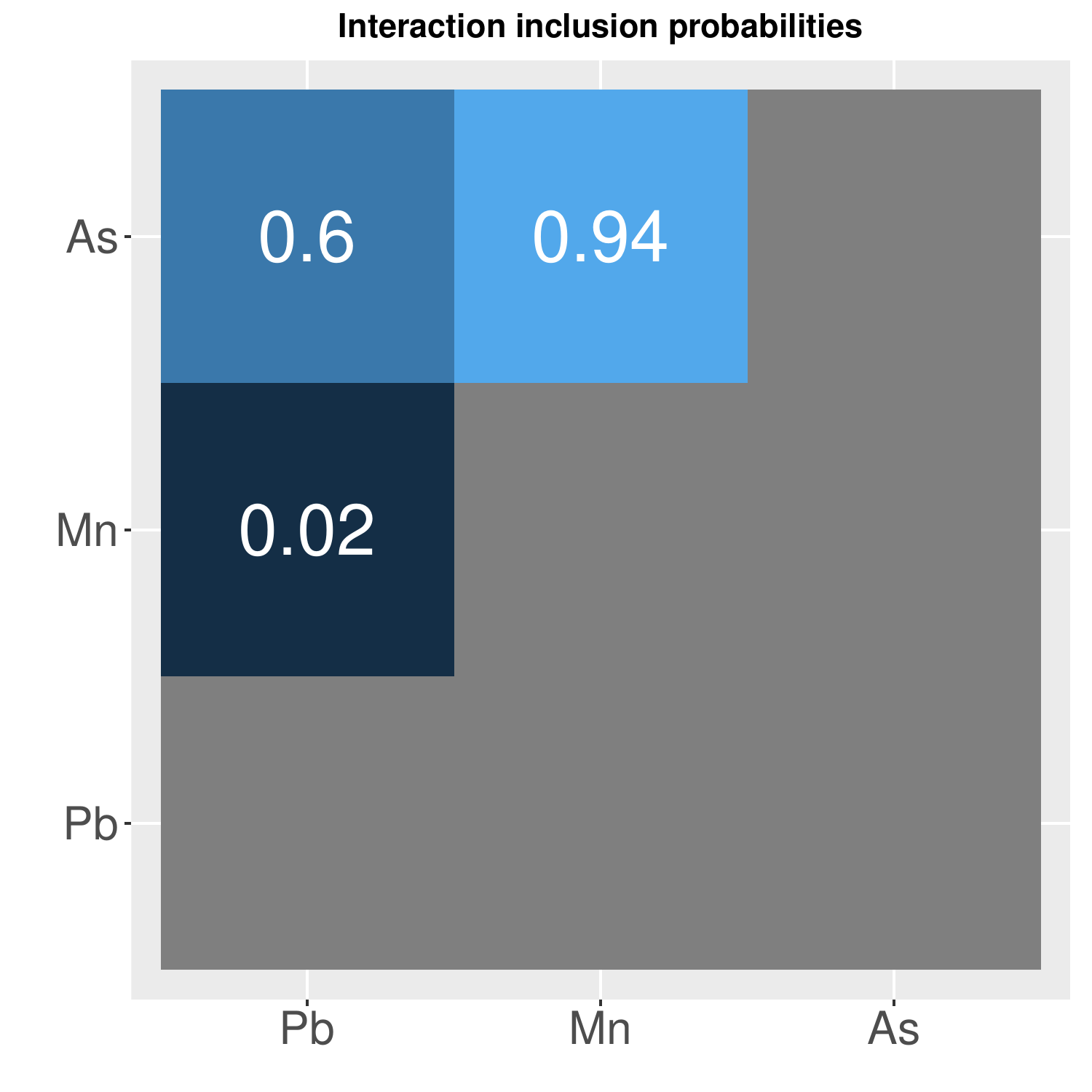} \hspace{0.4cm}
\includegraphics[height=2.5in]{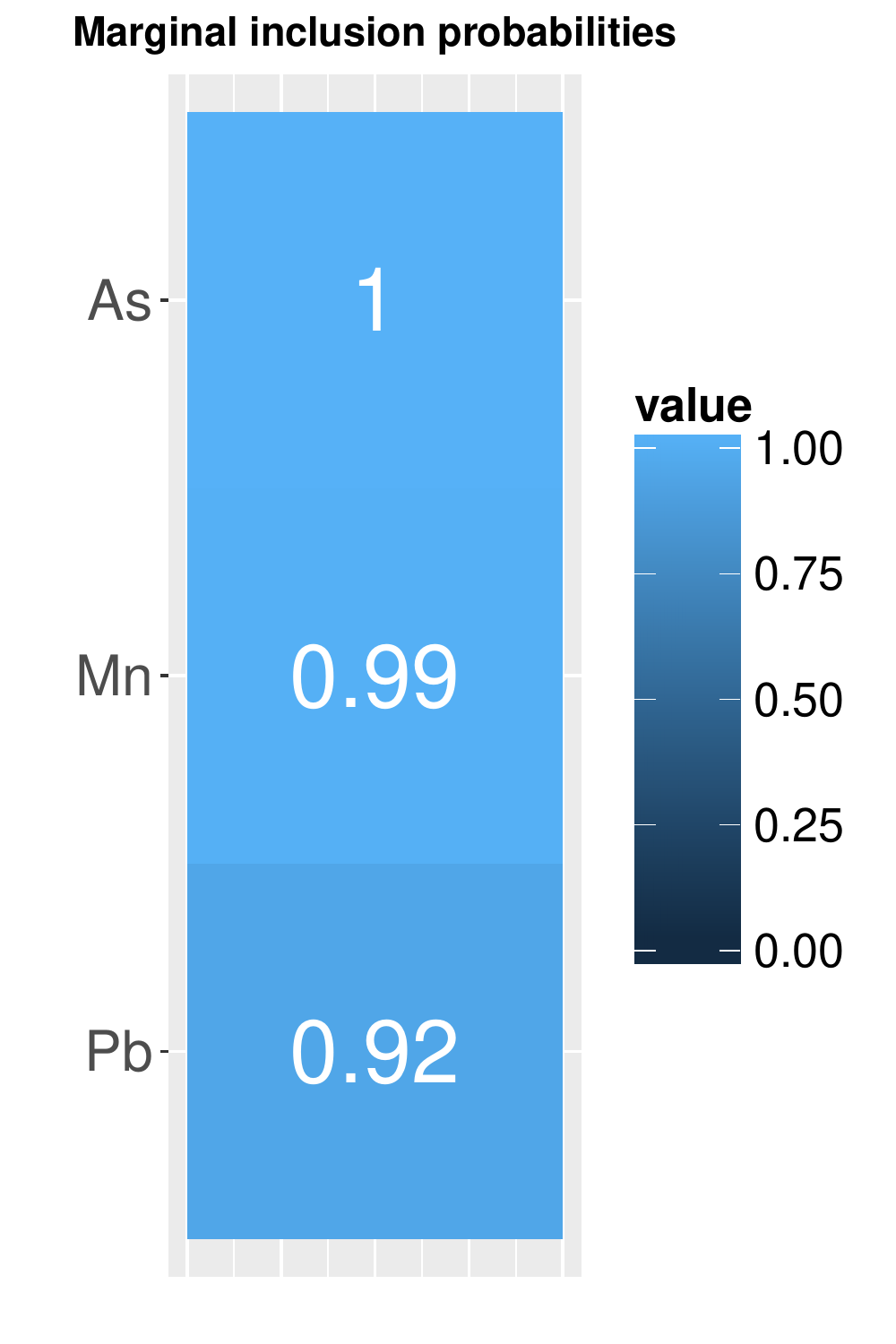}
\caption{The left panel contains posterior inclusion probabilities for two-way interactions, and the right panel shows the marginal inclusion probabilities for each exposure}
\label{fig:BangladeshProb}
\end{figure}

One feature of the proposed approach is the ability to examine the posterior distribution of $\boldsymbol{\zeta}$ in order to identify whether individual exposures are important for outcome prediction, whether there is evidence of an interaction between two metals or any higher-order interactions. Figure \ref{fig:BangladeshProb} shows the posterior probability of two-way interactions between all metals in the model, and the marginal posterior inclusion probability for each metal exposure. Figure \ref{fig:BangladeshProb} provides strong evidence that each exposure is associated with a child's MCS score. This confirms the preliminary results of Section \ref{sec:prelim} that showed associations between each of the three exposures and the MCS score. A key distinction of the results here and those seen in Section \ref{sec:prelim} is that it also provides strong evidence of an interaction between Mn and As, as well as moderate evidence of an interaction between Pb and As. Not shown in the figure is the posterior probability of a three-way interaction among Pb, Mn, and As, which was estimated to be 0.0005.

\subsection{Visualization of identified interactions}

\begin{figure}[ht]
\centering
	  \includegraphics[width=0.95\linewidth]{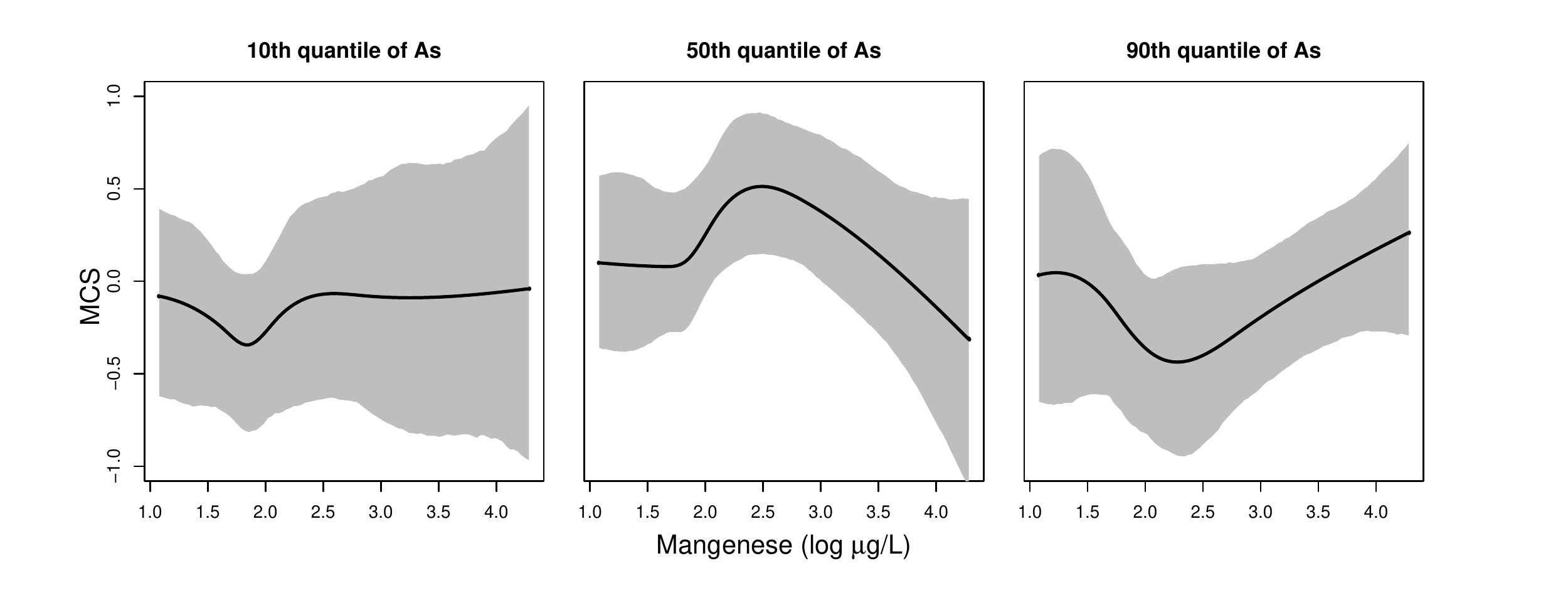}
\caption{Posterior means and pointwise 95\% credible intervals of the effect of Mangenese on MCS at different quantiles of Arsenic while fixing Lead at its median.}
\label{fig:twoway}
\end{figure}

Now that we have identified interactions in the model, we plot posterior predictive distributions to assess the form of these associations. Figure \ref{fig:twoway} plots the relationship between Mn and MCS for different values of As, while fixing Pb at its median. At the median level of As, we estimate an inverted U-shaped effect of Mn on MCS scores. This matches results seen previously in a kernel machine regression analysis of these data \cite{Bobb2014}, which provided graphical evidence of an interaction between Mn and As on the MCS score but didn't provide a measure of the strength of this interaction.

\begin{figure}[ht]
\centering
	  \includegraphics[width=0.95\linewidth]{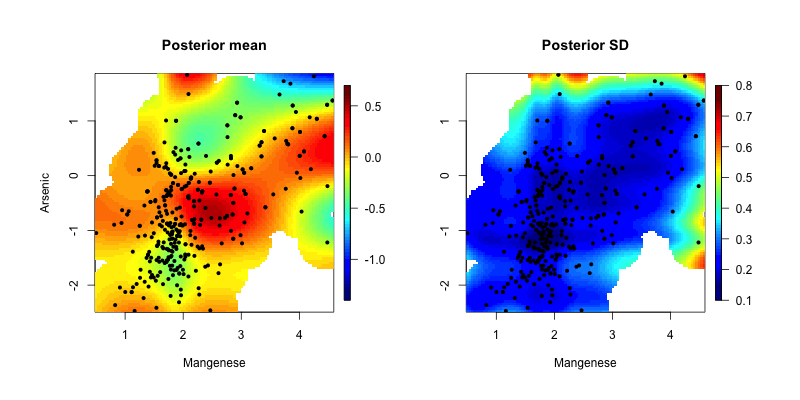}
\caption{Posterior means and standard deviation of the relationship between Mangenese and Arsenic on MCS. The black dots represent the location of the observed data points. }
\label{fig:heatmaps}
\end{figure}

We also examine surface plots that show the  exposure-response relationship between As, Mn, and MCS across the range of As and Mn values, not just specific quantiles. Figure \ref{fig:heatmaps} shows both the posterior mean and pointwise standard deviations of this relationship. We see that the median levels of Mn and As are the most beneficial towards children in terms of their MCS score, which corresponds to the hump in the middle panel of Figure \ref{fig:twoway}. This is a somewhat unexpected result given that As is not beneficial even at low levels, and could be due to residual confounding \citep{valeri2017joint}. The posterior standard deviations of this estimate indicate that the uncertainty around this surface is the smallest in the center, where most of the data is observed, and greater on the fringes of the observed data, where we are effectively extrapolating effects. 

\section{Discussion}
\label{sec:discussion}

We have proposed a flexible method for analyzing complex, nonlinear associations between multiple exposures and a continuous outcome. Our method improves upon existing methodology for environmental mixtures in several ways. By not relying on Gaussian processes to introduce flexibility in the exposure-response relationship,  our approach can handle larger data sets on the order of tens of thousands of subjects. The spike and slab formulation of our model also allows for formal inference on the presence of interactions among exposures, an important question in environmental epidemiology. Previous studies have used graphical tools and inference on summary statistics characterizing the effects of one pollutant given the levels of others to identify interactions between pollutants. While useful, these approaches do not provide uncertainty assessment of the interaction as a whole. Moreover, it can be difficult to visualize higher-order interactions. Once we have samples of the $\boldsymbol{\zeta}$ matrix, we can easily perform inference on more complex features of the model such as higher-order interactions and whether or not exposures are more likely to enter the model together or separately. We have illustrated the strengths of this approach through simulated data, and a study of the health effects of metal mixtures on child neurodevelopment in Bangladesh. 

A natural question that arises in the application of our method is how does it perform on higher-dimensional mixtures in real-world settings.  We evaluated a situation with both a large sample size and number of exposures in Section \ref{sec:simHD} of our simulation study, and found that it performs well at identifying higher-order interactions. To further illustrate this point, we have analyzed a secondary data set from the National Health and Nutrition Examination Survey (NHANES), which has a larger sample size and more exposures than the Bangladesh data. The results of this study can be found in Appendix A. 

Throughout the manuscript we have assumed that the relationship between the exposures and outcome is constant across levels of the covariate space $\boldsymbol{C}$. In some cases, there is interest in understanding if the exposure effect varies across levels of particular covariates such as age or gender, which is frequently referred to as treatment effect heterogeneity. This could easily be explored using our model if the covariates are continuous. The covariates of interest can be placed with the exposures inside of the main $f()$ function, and our approach will be able to identify if any of these covariates modify the exposure effect. Identifying treatment effect heterogeneity in the presence of categorical covariates is not immediately applicable within our modeling framework, but it provides a topic of future research.

In the modeling framework we focused on a high-dimensional $\boldsymbol{X}$ and assumed that the number of covariates in $\boldsymbol{C}$ was relatively small, meaning we do not need to impose any shrinkage on $\boldsymbol{\beta}_C$. If the number of covariates is large as well, then shrinkage or spike and slab priors would need to be adopted. If interest is in estimation of $f(\boldsymbol{X})$, this could lead to what is called regularization induced confounding (RIC, \cite{hahn2017bayesian, hahn2018regularization}). RIC occurs because shrinkage leads to biased estimates of the coefficients in $\boldsymbol{\beta}_C$, and this bias leads to confounded estimates of the exposure effects of interest. Future work would look to extend the ideas here to the more difficult setting of a high-dimensional $\boldsymbol{C}$ when RIC could affect inference of the effects of interest. This would require reducing the shrinkage of the elements in $\boldsymbol{C}$ that are most associated with the elements of $\boldsymbol{X}$ in a manner that removes RIC bias. 

While our formulation enables the analysis of much larger data sets, there are some disadvantages relative to fully nonparametric techniques such as Gaussian processes. Our model requires us to specify the functional form of the relationships between the predictors and the outcome a priori. In the current implementation, we employed natural splines to flexibly model these relationships. While other formulations are possible, any formulation would require a subjective choice. Further, one must select the number of degrees of freedom of the splines, and we have restricted this number to be the same for all exposures across all features. This can be a limitation if some of the functions to be estimated are quite smooth while others are more complicated. Our approach is also not fully Bayesian, as we use the WAIC to select the degrees of freedom. However, our simulations show the proposed estimation procedure yields uncertainty estimates that are accurate in all of the scenarios we considered. Future extensions could consider a data-adaptive approach to the selection of the degrees of freedom for each function, or the utilization of other flexible structures that are both computationally fast and do not require pre-specification of the functional forms of the exposure-response functions.

\section*{Software} An R package implementing the proposed methodology is available at \\\url{github.com/jantonelli111/NLinteraction}

\section*{Acknowledgements}
This publication was made possible by USEPA grant RD-835872-01.  Its contents are solely the responsibility of the grantee and do not necessarily represent the official views of the USEPA.  Further, USEPA does not endorse the purchase of any commercial products or services mentioned in the publication. This work was also supported by grants from the National Institutes of Health (ES028811, ES000002, ES024332, ES007142, ES016454).

\appendix

\section{Effects of pollutants on telomere length in NHANES}

We will utilize data from the 2001-2002 cycle of the National Health and Nutrition Examination Survey (NHANES). The data is a representative sample of the United States and consists of 1003 adults. The outcome of interest is leukocyte telomere length (LTL). Telomeres are segments of DNA at the ends of chromosomes that are used to help protect chromosomes, and their lengths generally decrease with increasing age. In particular, LTL has been used as a proxy for overall telomere length, and has been shown to be associated with a number of adverse health outcomes \citep{mitro2015cross}. In the NHANES data, we have measurements from 18 persistent organic pollutants, which consist of 11 polychlorinated biphenyls (PCBs), 3 dioxins, and 4 Furans. It is believed that these environmental exposures could impact telomere length, though it is unclear how telomere length is associated with this particular mixture of exposures. Also in the data are additional covariates for which we need to adjust including age, sex, BMI, education status, race, lymphocyte count, monocyte count, cotinine level, basophil count, eosinophil count, and neutrophil count. We will analyze these data to assess whether any of the aforementioned environmental exposures are associated with LTL, and to search for any interactions between the environmental exposures. This quickly becomes a high-dimensional statistical issue even when looking at two-way interactions only as there are 153 two-way interactions alone. Therefore, this data requires a statistical approach to estimate high-dimensional models that allow for interactions and nonlinearities in the associations between the environmental mixture and LTL. 

\begin{figure}[ht]
\centering
	  \includegraphics[width=0.9\linewidth]{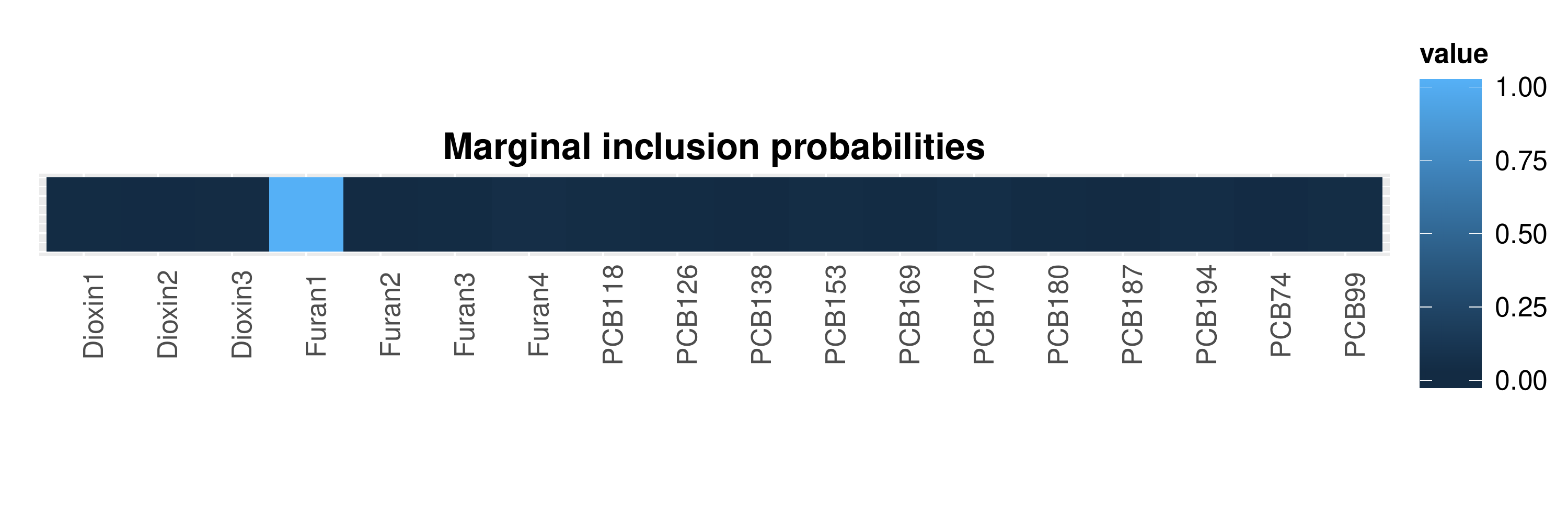}
\caption{Marginal posterior inclusion probabilities for each of the 18 exposures in the NHANES data.}
\label{fig:NHANESmarginal}
\end{figure}

We ran the Markov chain for 40,000 iterations with $M = 3, \gamma = p, a_0=0.001$, $b_0=0.001$, and a non-informative, normal prior on $\boldsymbol{\beta_c}$. We applied our approach with degrees of freedom, $d \in \{1,2,3,4,5,6,7 \}$ and we will present the results for $d=3$ as it was the model that minimized the WAIC. We examined the convergence of our MCMC using both trace plots and the potential scale reduction factor (PSR, \cite{gelman2014bayesian}) in Appendix H, and find that our model has converged. Figure \ref{fig:NHANESmarginal} shows the marginal inclusion probabilities for each of the 18 exposures. We see that all of the exposures have a posterior inclusion probability at or near zero with the exception of Furan1, which was included in the model 99\% of the time. We do not detect any higher-order interactions, as the posterior inclusion probabilities were exactly zero for all two-way and higher-order interactions. This provides strong evidence of a relationship between Furan1 and telomere length.

\begin{figure}[ht]
\centering
	  \includegraphics[width=0.5\linewidth]{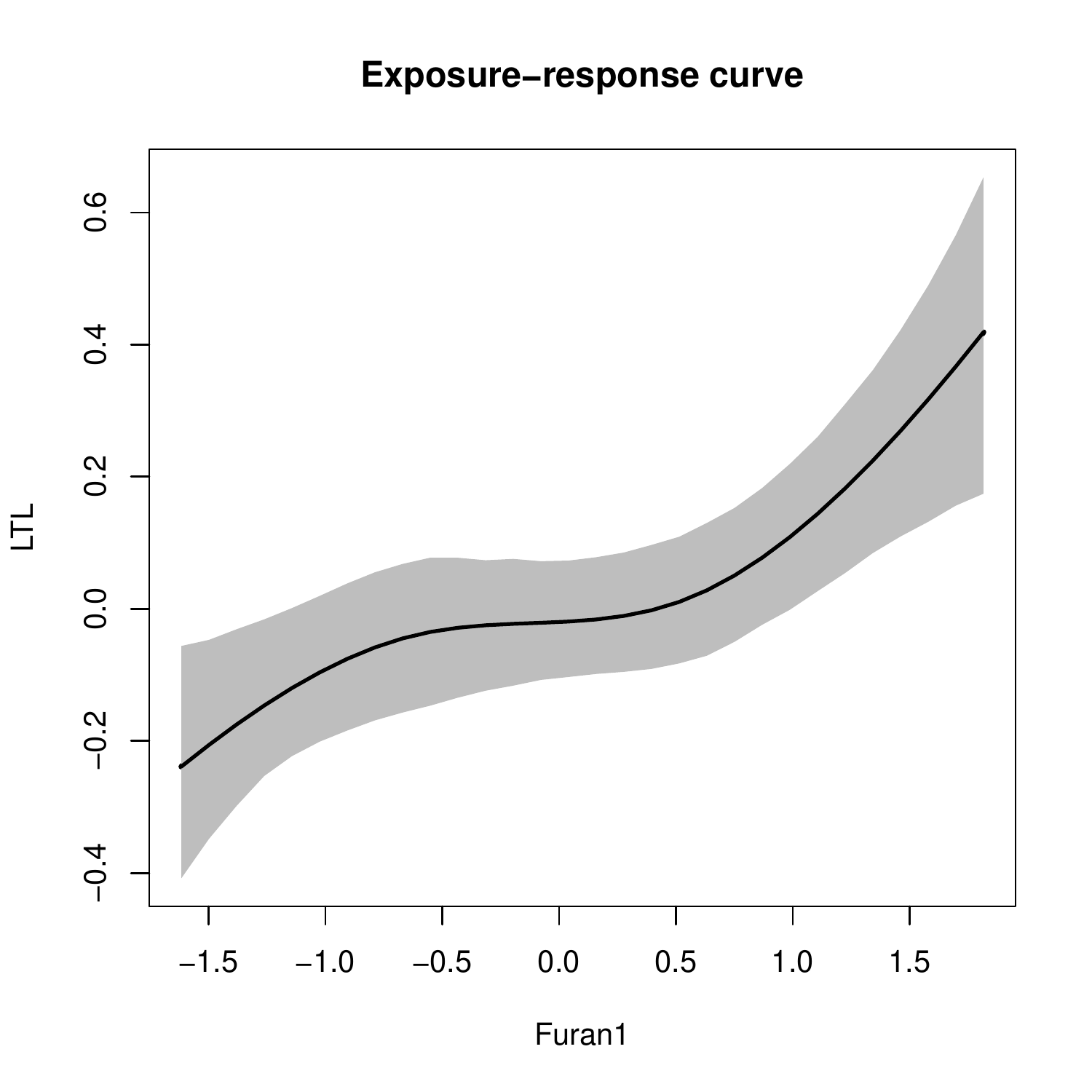}
\caption{The relationship between Furan1 and telomere length, while fixing all the other exposures at their medians.}
\label{fig:furan}
\end{figure}

Figure \ref{fig:furan} shows the effect of Furan1 on telomere levels while fixing other exposures at their medians. It is important to note that fixing other exposures at their medians does not change the shape of the curve in this instance, because there are no interactions between Furan1 and other exposures. We estimate a positive association between Furan1 exposure and telomere length. The curve is somewhat nonlinear as seen in both figure \ref{fig:furan} and the fact that the WAIC chose three degrees of freedom for the basis functions. 

\section{Simulation details and additional simulation results}

Here we present more details regarding the simulation study in Section 5.1 of the manuscript, an additional simulation study with increased correlation between the exposures and covariates, as well as an additional simulation study that does not include any interactions, but has nonlinear main effect terms. 

\subsection{Details of data generation for simulation 5.1}

In Section 5.1 of the manuscript, we generated exposures from a multivariate normal distribution with mean $\boldsymbol{\mu}_X = \boldsymbol{C \alpha}$, and then generated the outcome from the following model:
\begin{align*}
	Y = 0.7 X_2 X_3 + 0.6 X_4^2 X_5 + \boldsymbol{C \beta}_C + \epsilon
\end{align*}

We used the values $ \boldsymbol{\beta}_C = [0.2, 0.3, 0, 0.4, -0.2, -0.3, -0.1, -0.5, 0.3, 0.25]$ for the outcome, and the following values for $\boldsymbol{\alpha}$:
$$
\begin{pmatrix}
	0.2 & 0 & 0 & 0 & 0 & 0 & 0 & 0 & 0 & 0.1\\
	0.3 & -0.2 & 0 & 0 & 0 & 0 & 0 & 0 & 0 & 0\\
	0 & 0.4 & 0.2 & 0 & 0 & 0 & 0 & 0 & 0 & 0\\
	0 & 0 & -0.3 & 0.1 & 0 & 0 & 0 & 0 & 0 & 0\\
	0 & 0 & 0 & 0.6 & 0.4 & 0 & 0 & 0 & 0 & 0\\
	0 & 0 & 0 & 0 & -0.1 & -0.3 & 0 & 0 & 0 & 0\\
	0 & 0 & 0 & 0 & 0 & -0.3 & -0.15 & 0 & 0 & 0\\
	0 & 0 & 0 & 0 & 0 & 0 & 0.45 & 0.1 & 0 & 0\\
	0 & 0 & 0 & 0 & 0 & 0 & 0 & 0.5 & -0.2 & 0\\
	0 & 0 & 0 & 0 & 0 & 0 & 0 & 0 & -0.2 & 0.2
\end{pmatrix}
$$

\begin{figure}[ht]
\centering
	  \includegraphics[width=0.6\textwidth]{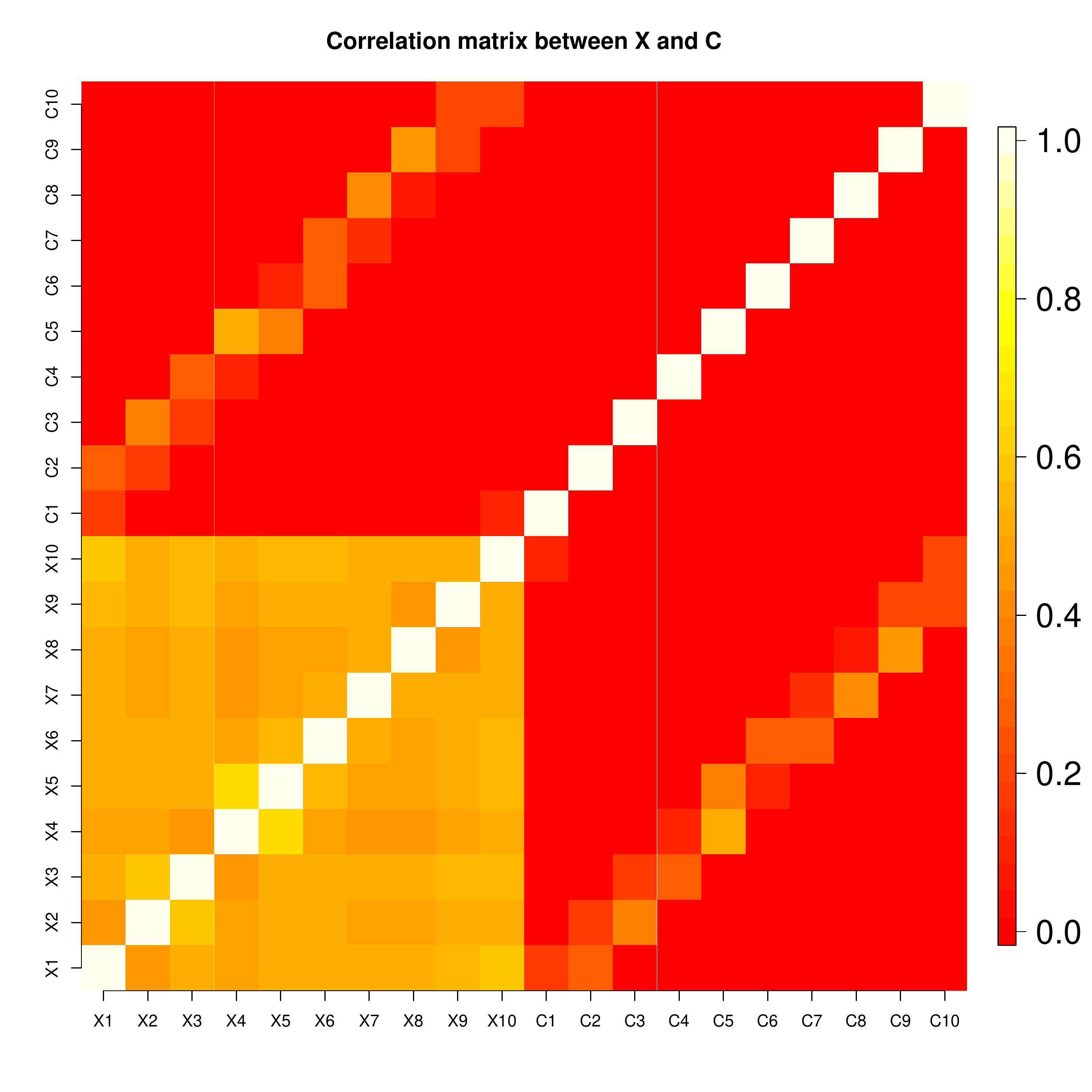} 
\caption{Correlation matrix among all exposures and covariates for simulation 5.1}
\label{fig:corXC}
\end{figure}

We can also look at the correlation matrix between $\boldsymbol{X}$ and $\boldsymbol{C}$ to understand the impact of these parameter choices. We can see in Figure \ref{fig:corXC} that there is a very high amount of correlation between the 10 elements of $\boldsymbol{X}$ denoted by the lower left block in the correlation matrix. Looking at the correlation between $\boldsymbol{X}$ and $\boldsymbol{C}$ we can see that each element of $\boldsymbol{X}$ has certain elements of $\boldsymbol{C}$ that are correlated with it.  

\subsection{Increasing correlation between X and C}

\begin{figure}[ht]
\centering
	  \includegraphics[width=0.6\textwidth]{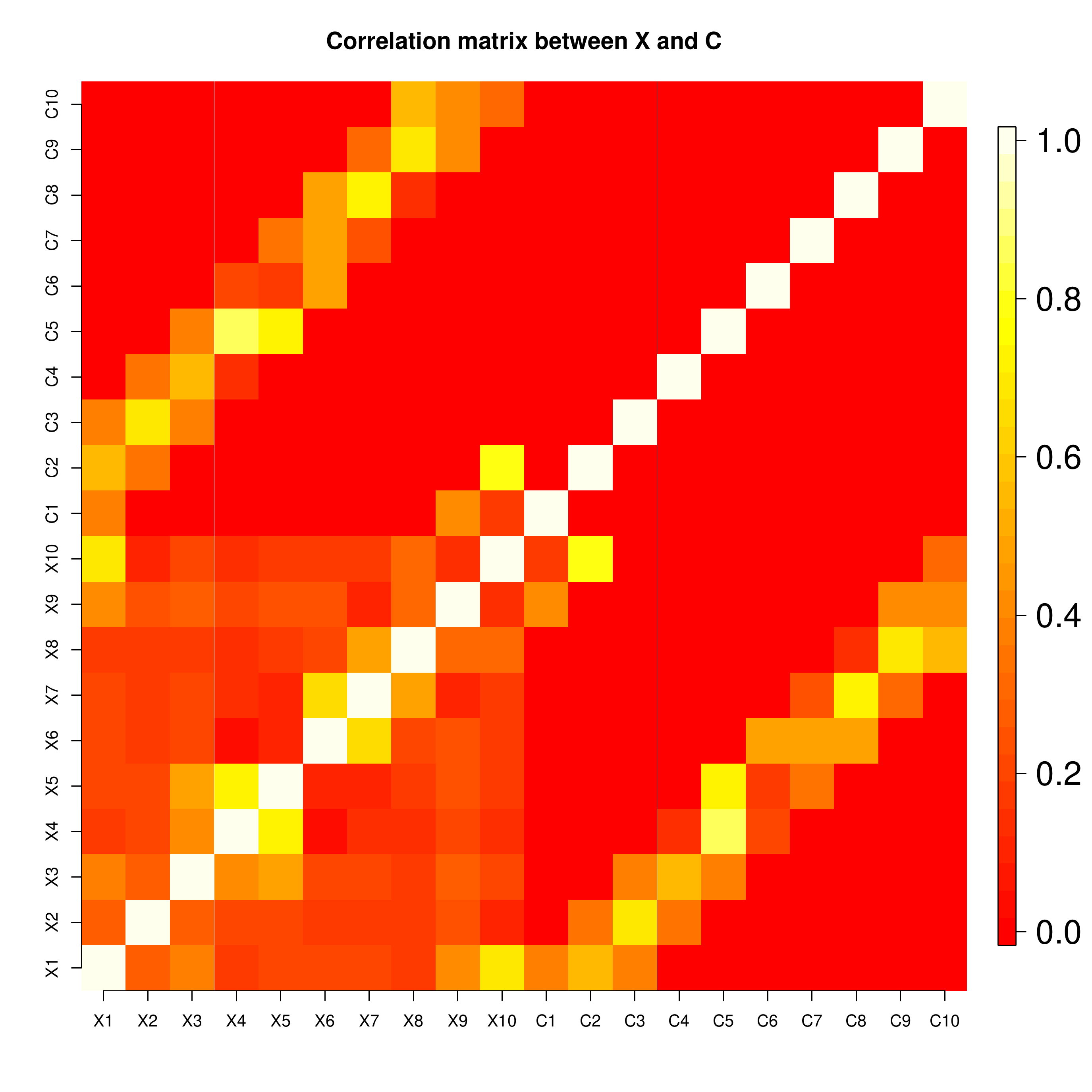} 
\caption{Correlation matrix among all the exposures and covariates for the simulation with increased correlation between covariates and exposures}
\label{fig:corXCincrease}
\end{figure}

Now we will increase the correlation between $\boldsymbol{X}$ and $\boldsymbol{C}$ to see how inference is impacted by this correlation. We will now let the mean of each exposure be a function of three covariates (instead of 2 as in simulation 5.1) and we will increase the magnitude of this correlation. The correlation matrix for this new setup can be found in Figure \ref{fig:corXCincrease}, and we can see there is substantially more correlation between $\boldsymbol{X}$ and $\boldsymbol{C}$. We will compare results from two situations: the results from simulation 5.1 of the manuscript,  and this new simulation with increased correlation between the exposures and covariates. The data generating models for the outcome are the same between both these scenarios with the main difference being the correlation matrices for $\boldsymbol{X}$ and $\boldsymbol{C}$. 

\begin{figure}[ht]
\centering
	  \includegraphics[width=0.4\textwidth]{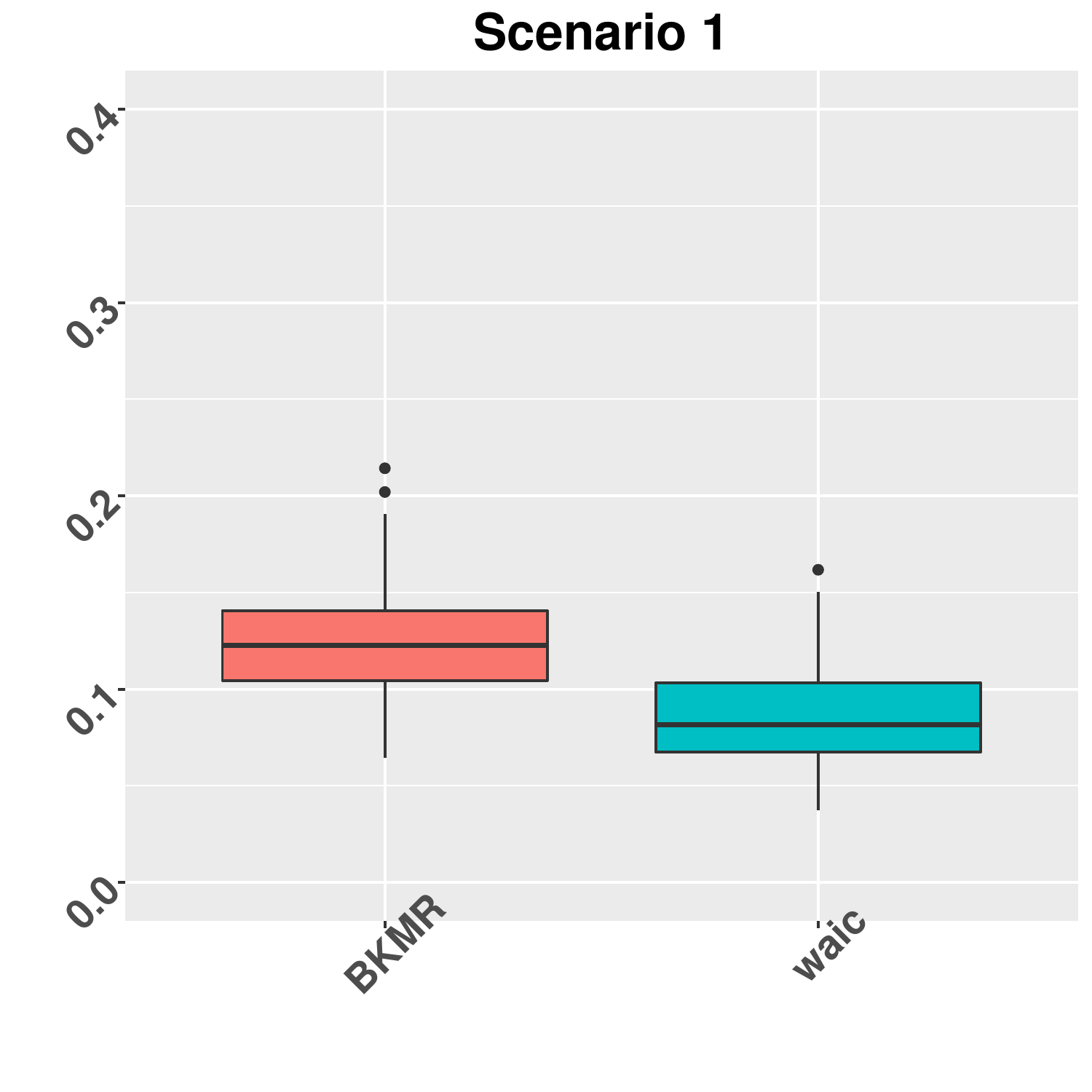}
	  \includegraphics[width=0.4\textwidth]{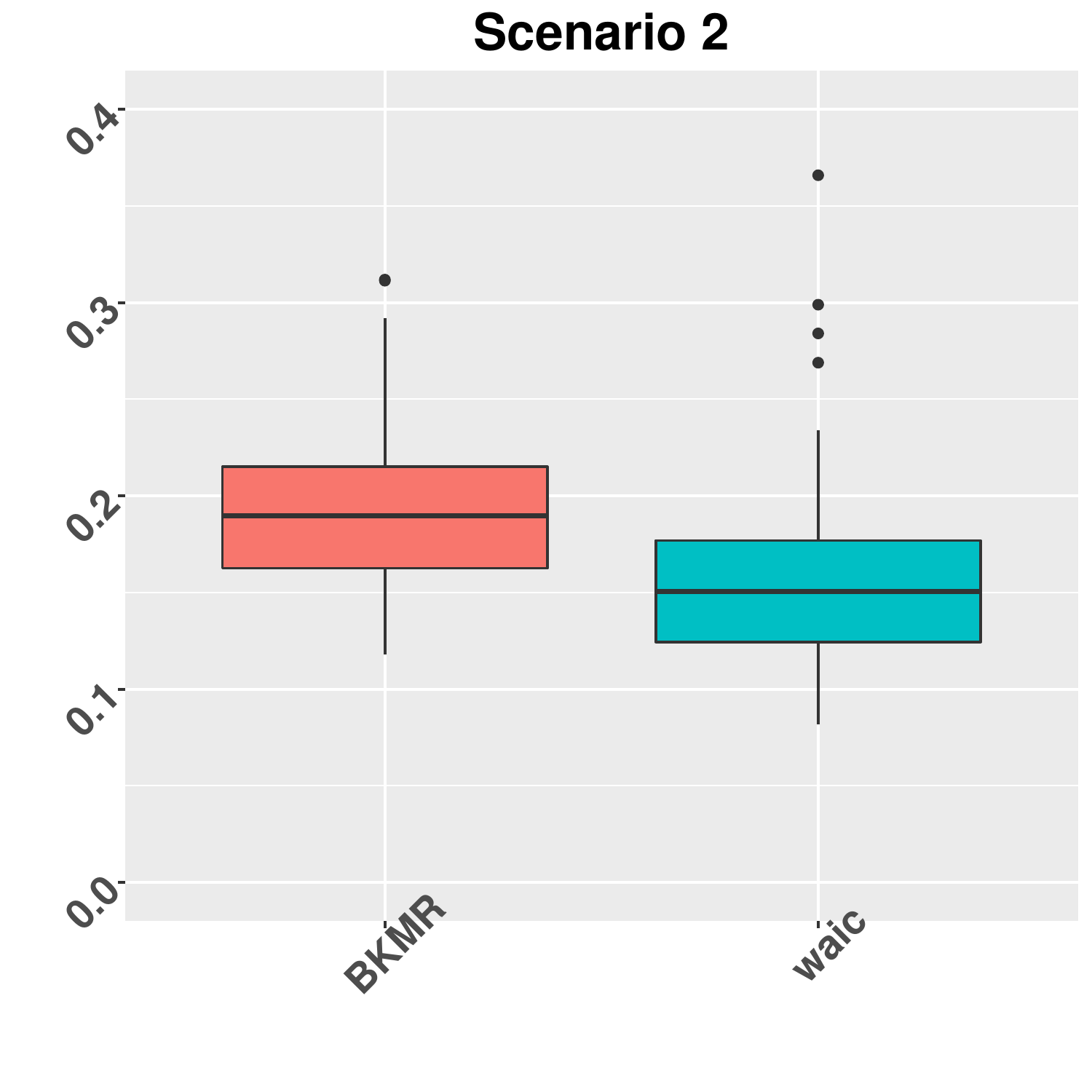} 
\caption{Mean squared error in estimating $f(X_i)$ for the two approaches considered under both simulation scenarios. The first scenario is that of simulation 5.1 of the manuscript, and the second simulation is with the increased correlation between the covariates and exposures.}
\label{fig:simCorXCmse}
\end{figure}

The mean squared error from this additional scenario can be seen in Figure \ref{fig:simCorXCmse}. We only highlight our approach based on WAIC to choose the degrees of freedom and the BKMR approach as these two were the best performing approaches in the manuscript. We see that the performance of both approaches degrades when the correlation is increased between the covariates and exposure.  This increase in MSE is fairly similar between the two approaches, however, and the approach based on WAIC performs the best between the two approaches. Not shown in the figure is that the approach based on WAIC also achieves interval coverages of 91\% indicating that performance of the approach is still good even under high levels of correlation between the covariates and exposures. 

\subsection{Additional simulation study}

Now we present a simulation study that is similar to the one presented in Section 5.1 of the manuscript, except the effect of the exposures has been changed. The covariates and exposures are generated from the same multivariate normal distribution as before. Now the outcome is generated from the following model:

\begin{align*}
	Y = 0.8 \text{sin}(\pi X_1) + 0.4 (X_3^2 - 0.5) + 0.2 \text{exp}(X_4) + 0.6 X_5 + \boldsymbol{C \beta}_C +  \epsilon
\end{align*}

\begin{figure}[ht]
\centering
	  \includegraphics[height=0.255\textheight]{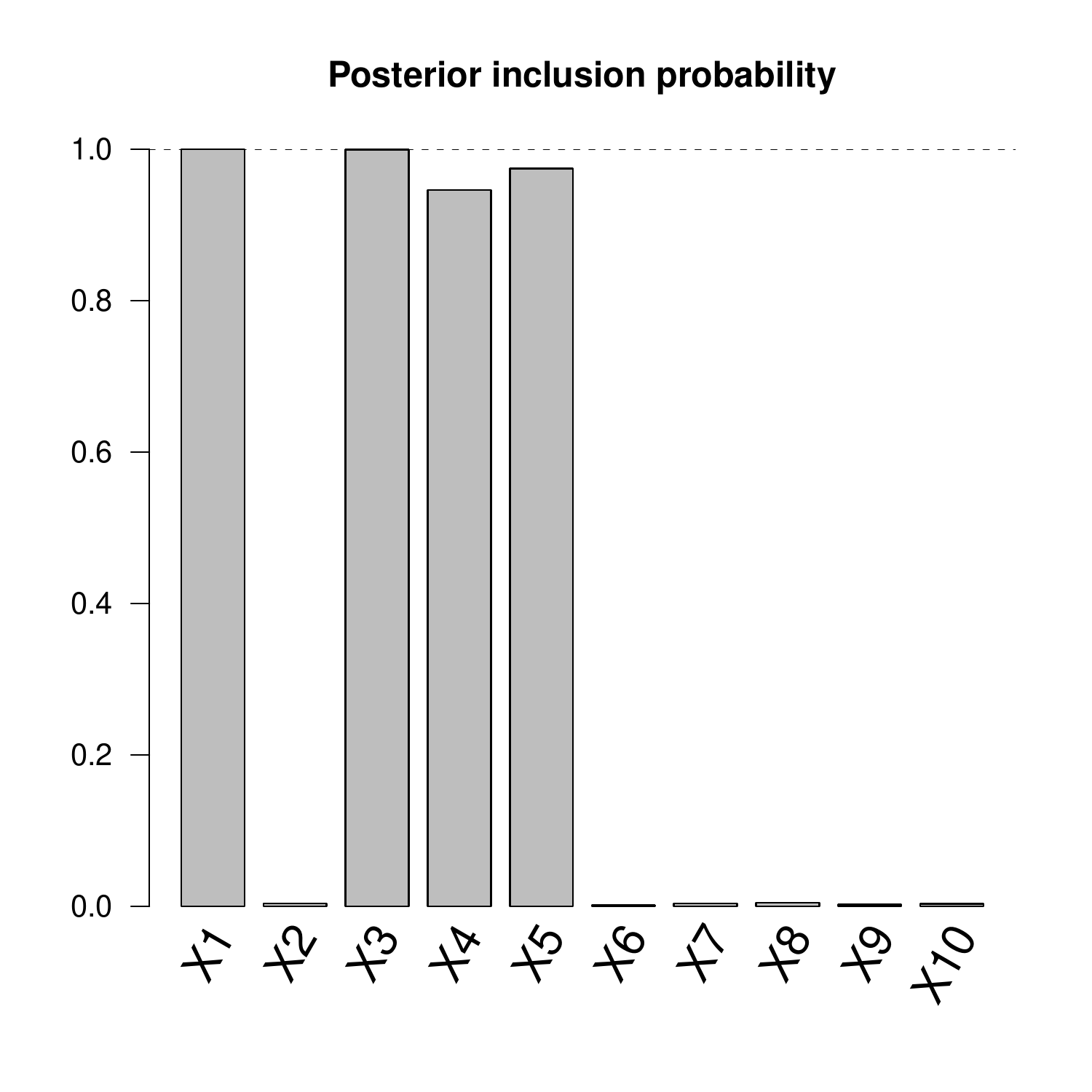} 
      \includegraphics[height=0.245\textheight]{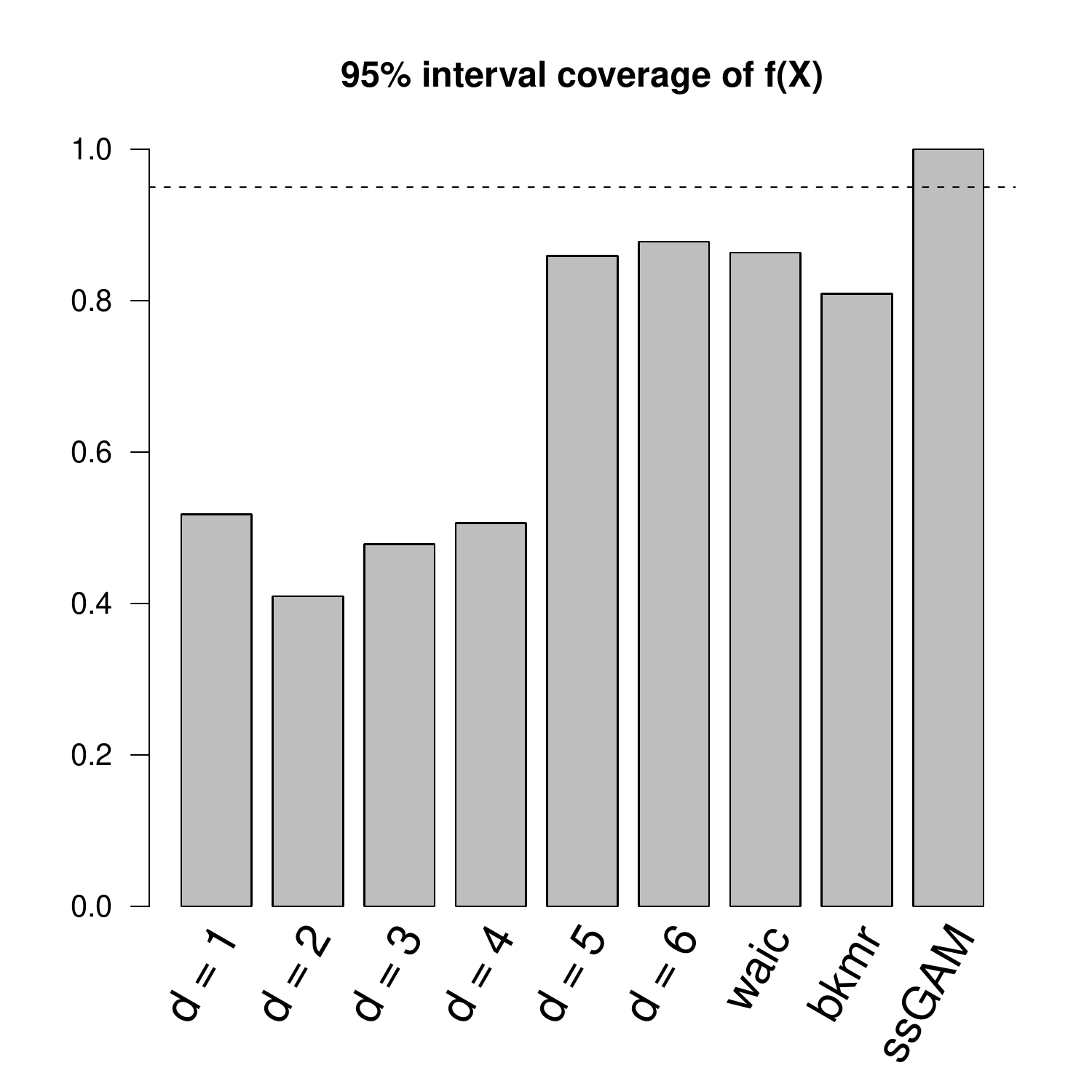} \\
      \includegraphics[height=0.25\textheight]{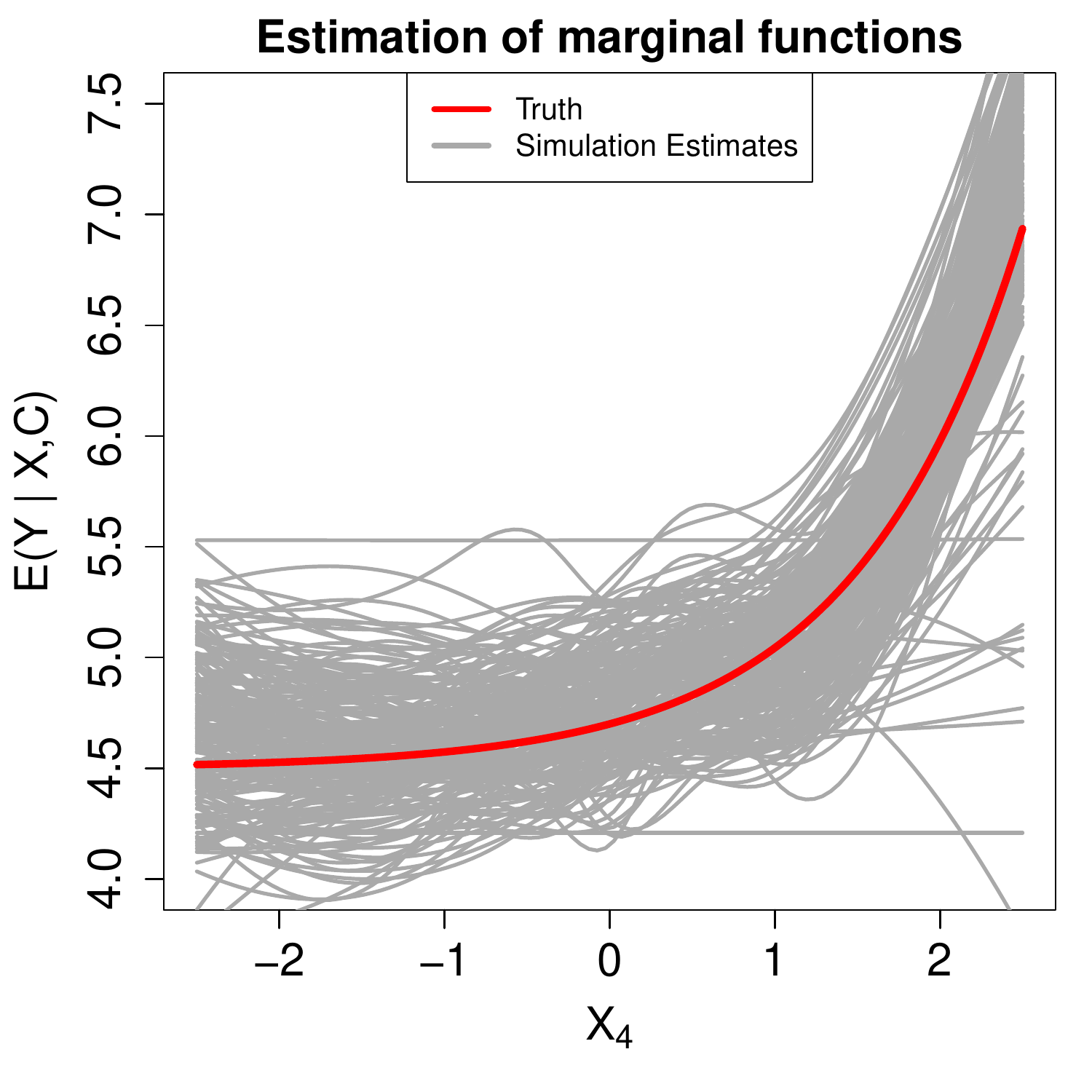}
      \includegraphics[height=0.247\textheight]{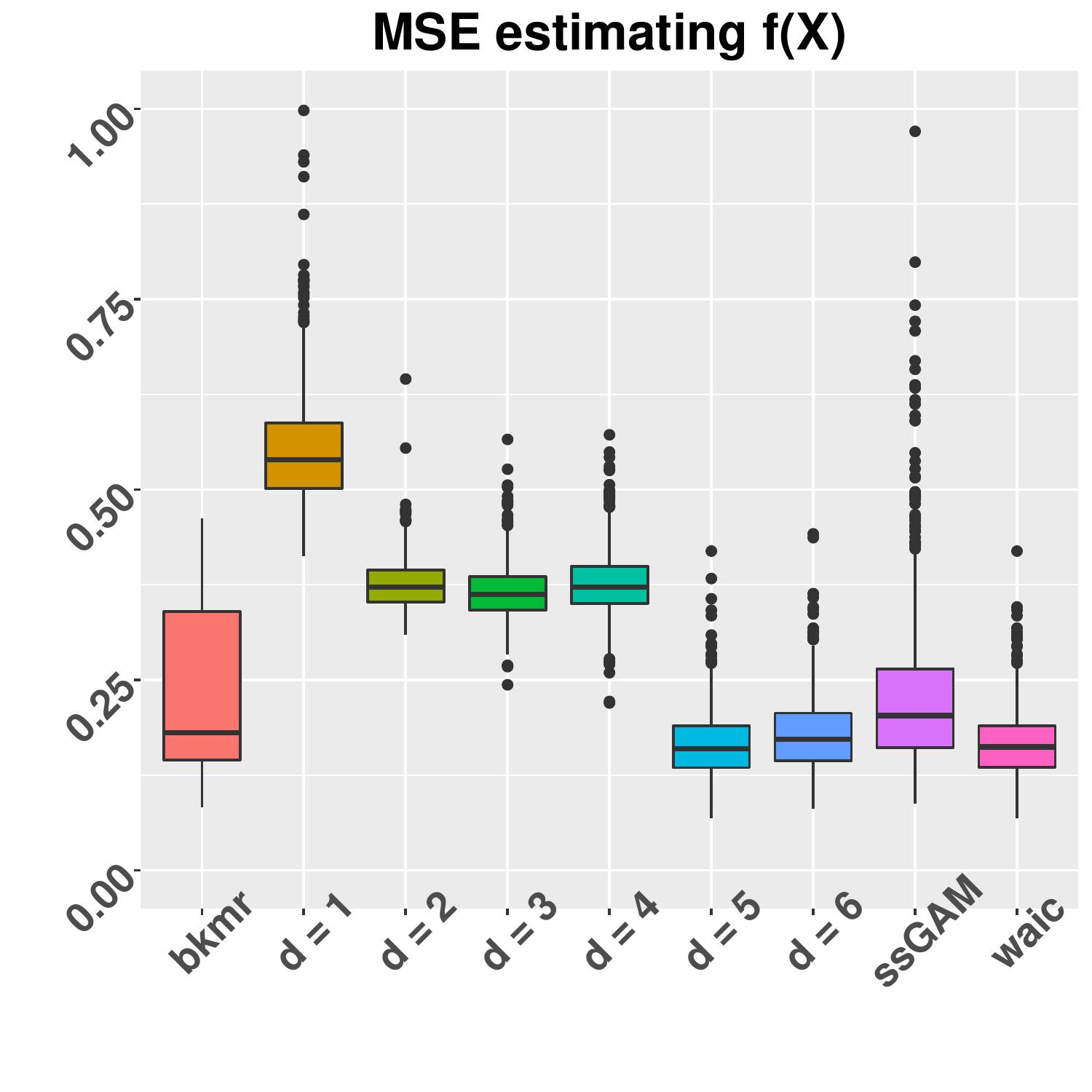}
\caption{Results of the additional simulation study. The top left panel shows the posterior inclusion probabilities for each exposure for the model selected by the WAIC. The top right panel shows the 95\% credible interval coverages of $f(X_i)$. The bottom left panel shows the estimated marginal functions of $X_4$ on Y for the model selected by the WAIC, where the red line represents the truth and the grey lines represent individual estimates from separate simulations. The bottom right panel shows the MSE in estimating $f(X_i)$ for all models considered.}
\label{fig:simadd1}
\end{figure}

The results can be seen in Figure \ref{fig:simadd1}. The top left panel shows that our model is correctly identifying the four important exposures while excluding the others. Not shown in the figure is that all two-way interactions had posterior inclusion probabilities less than 0.01, indicating that our model correctly removes all higher-order interactions. The top right panel shows that none of the approaches are able to achieve the nominal coverage in this instance, though again ssGAM has confidence intervals that are far too wide leading to a coverage probability of one. Of the other estimators though, our approach with degrees of freedom chosen by WAIC performs best in terms of coverage probability with 86\%. The bottom left panel shows that we are correctly capturing the nonlinear relationship between $X_4$ and $Y$. Lastly, the bottom right panel shows that our approach obtains the lowest MSE of all approaches considered.

\section{Posterior computation}

Samples from the posterior distributions of all unknown parameters in Model 3.2 can easily be obtained via Gibbs sampling, as all of the full conditional distributions have closed form. The algorithm is as follows: 

\begin{enumerate}
	\item Sample $\sigma^2$ from the following full conditional:
	\begin{align*}
		\sigma^2 \vert \bullet \sim IG \left(a^*, b^*  \right),
	\end{align*}
	where
	$$a^* = a_0 + \frac{n}{2} + \frac{|\boldsymbol{\beta}|_0}{2},$$
	and
	$$b^* = b_0 + \frac{\sum_{i=1}^n \left( Y_i - \sum_{h=1}^k  f_{h}(\boldsymbol{X_i}) - \boldsymbol{C_i\beta_c} \right)^2}{2}  + \sum_{S:\beta_S \neq 0} \frac{(\beta_S - \mu_{\beta_S})^2}{2\sigma_{\boldsymbol{\beta}}^2}$$
	\item Sample $\tau_h$ for $h=1,\dots,k$ from the following full conditional:
	\begin{align*}
		\tau_h \vert \bullet \sim Beta \left(M + \sum_{j=1}^p \zeta_{jh}, \gamma + \sum_{j=1}^p (1 - \zeta_{jh}) \right)
	\end{align*}
    \item To update $\boldsymbol{\zeta}$ let us first introduce some additional notation. Let $\boldsymbol{\zeta_{B}}$ be the subset of $\boldsymbol{\zeta}$ we are updating and $\boldsymbol{\beta_{B}}$ be the corresponding regression coefficients. Further, let $\boldsymbol{\Lambda}$ be the set of all parameters in our model excluding $\boldsymbol{\zeta_{B}}$ and $\boldsymbol{\beta_{B}}$. To update $\boldsymbol{\zeta_{B}}$ we need to calculate $p(\boldsymbol{\zeta_{B}} \vert \boldsymbol{D}, \boldsymbol{\Lambda})$, which we can do via the following:
    \begin{align}
    	\label{eqn:zetaupdate} \nonumber p(\boldsymbol{\zeta_{B}} = \boldsymbol{z_B} \vert \boldsymbol{D}, \boldsymbol{\Lambda}) &= \frac{p(\boldsymbol{\beta_{B}} = \boldsymbol{0}, \boldsymbol{\zeta_B} = \boldsymbol{z_B} \vert \boldsymbol{D}, \boldsymbol{\Lambda})}{p(\boldsymbol{\beta_{B}} = \boldsymbol{0} \vert \boldsymbol{\zeta_B} = \boldsymbol{z_B},  \boldsymbol{D}, \boldsymbol{\Lambda})} \\
	\nonumber &= \frac{p( \boldsymbol{D}, \boldsymbol{\Lambda} \vert \boldsymbol{\beta_{B}} = \boldsymbol{0}, \boldsymbol{\zeta_B} = \boldsymbol{z_B}) p(\boldsymbol{\beta_{B}} = \boldsymbol{0}, \boldsymbol{\zeta_B} = \boldsymbol{z_B})}{p( \boldsymbol{D}, \boldsymbol{\Lambda}) p(\boldsymbol{\beta_{B}} = \boldsymbol{0} \vert \boldsymbol{\zeta_B} = \boldsymbol{z_B},  \boldsymbol{D}, \boldsymbol{\Lambda})} \\
\nonumber &= \frac{p( \boldsymbol{D}, \boldsymbol{\Lambda} \vert \boldsymbol{\beta_{B}} = \boldsymbol{0}) p(\boldsymbol{\beta_{B}} = \boldsymbol{0}, \boldsymbol{\zeta_B} = \boldsymbol{z_B})}{p( \boldsymbol{D}, \boldsymbol{\Lambda}) p(\boldsymbol{\beta_{B}} = \boldsymbol{0} \vert \boldsymbol{\zeta_B} = \boldsymbol{z_B},  \boldsymbol{D}, \boldsymbol{\Lambda})} \\
\nonumber & \propto \frac{p(\boldsymbol{\beta_{B}} = \boldsymbol{0}, \boldsymbol{\zeta_B} = \boldsymbol{z_B})}{p(\boldsymbol{\beta_{B}} = \boldsymbol{0} \vert \boldsymbol{\zeta_B} = \boldsymbol{z_B},  \boldsymbol{D}, \boldsymbol{\Lambda})} \\
&=  \frac{\Phi(\boldsymbol{0}; \boldsymbol{\mu_{\beta}}, \boldsymbol{\Sigma_{\beta}})}{\Phi(\boldsymbol{0}; \boldsymbol{M}, \boldsymbol{V})} \prod_{h,j: \zeta_{jh} \in \boldsymbol{\zeta_B}} \tau_h^{\zeta_{jh}} (1 - \tau_h)^{1 - \zeta_{jh}} 
    \end{align}
where $\Phi()$ represents the multivariate normal density function. $\boldsymbol{M}$ and $\boldsymbol{V}$ represent the conditional posterior mean and variance for the elements of $\boldsymbol{\beta}$ corresponding to the elements of $\boldsymbol{\zeta_B}$ that are equal to one and are defined as 
    \begin{align*}
    	\boldsymbol{M} = \left(\frac{\boldsymbol{{X^*}}^T \boldsymbol{{X^*}}}{\sigma^2} + \boldsymbol{\Sigma_{\beta}}^{-1} \right)^{-1} \bigg(\frac{{\boldsymbol{{X^*}}}^T Y^*}{\sigma^2} + \boldsymbol{\Sigma_{\beta}}^{-1} \boldsymbol{\mu_{\beta}}\bigg)
	\end{align*}
	\begin{align*}
	\boldsymbol{V} = \left(\frac{\boldsymbol{{X^*}}^T \boldsymbol{{X^*}}}{\sigma^2} + \boldsymbol{\Sigma_{\beta}}^{-1} \right)^{-1},
    \end{align*}
	where $\boldsymbol{X^*}$ is a design matrix with the appropriate main effects and interactions induced by the model $\boldsymbol{z_B}$, and ${Y_i}^* = Y_i - \sum_{h: h\not\in \boldsymbol{\zeta_B}} f_h(\boldsymbol{X}) - \boldsymbol{C_i\beta_c}$.
    Intuitively, the numerator in (\ref{eqn:zetaupdate}) is the prior probability that all elements of $\boldsymbol{\beta}$ corresponding to features in $\boldsymbol{\zeta_B}$ are zero and $\boldsymbol{\zeta_B} = \boldsymbol{z_B}$. The denominator in (\ref{eqn:zetaupdate}) represents the conditional posterior probability that all elements of $\boldsymbol{\beta}$ corresponding to features in $\boldsymbol{\zeta_B}$ are zero. 
	\item Update $\boldsymbol{\beta_{S}^{(h)}}$ from a normal distribution with mean $\boldsymbol{M}$ and variance $\boldsymbol{V}$ as defined above. 
        \item Update $\boldsymbol{\beta_c}$ from the following distribution:
    \begin{align*}
    	\text{MVN} \left(\left(\frac{\boldsymbol{C}^T \boldsymbol{C}}{\sigma^2} + \boldsymbol{\Sigma_c}^{-1} \right)^{-1} (\frac{\boldsymbol{C}^T \widetilde{Y}}{\sigma^2} + \boldsymbol{\Sigma_c^{-1} \mu_c}),\left(\frac{\boldsymbol{C}^T \boldsymbol{C}}{\sigma^2} + \boldsymbol{\Sigma_c^{-1}} \right)^{-1} \right),
    \end{align*}
    where $\widetilde{Y_i} = Y_i - \sum_{h=1}^{k} f_{h}(\boldsymbol{X_i})$
\end{enumerate}

\section{Derivation of empirical Bayes variance}

As seen in \cite{casella2001empirical}, the Monte Carlo EM algorithm treats the parameters as missing data and iterates between finding the expectation of the ``complete data'' log likelihood, where expectations are calculated from draws of a gibbs sampler, and maximizing this expression as a function of the hyperparameter values. After removing the terms not involving $\sigma_{\boldsymbol{\beta}}$, we write the ``complete data'' log likelihood  as

\begin{align*}
	-\frac{|\boldsymbol{\beta}|_0}{2} \log (\sigma_{\boldsymbol{\beta}}^2) - \sum_{S:\beta_S \neq 0} \frac{(\beta_S - \mu_{\beta_S})^2}{2\sigma^2  \sigma_{\boldsymbol{\beta}}^2},
\end{align*}

where $\mu_{\beta_S}$ is the prior mean for coefficient $\beta_S$, and $|\boldsymbol{\beta}|_0$ is the number of nonzero regression parameters. The estimate of $\sigma_{\boldsymbol{\beta}}^2$ at iteration $m$ uses the posterior samples from iteration $m-1$, which were sampled with $\sigma_{\boldsymbol{\beta}}^2 = {\sigma_{\boldsymbol{\beta}}^2}^{(m-1)}$. Adopting the same notation that is typically used with the EM algorithm, we compute the expectation of this expression, where expectations are taken as the averages over the previous iterations posterior samples: 

\begin{align*}
	Q(\sigma_{\boldsymbol{\beta}}^2 | {\sigma_{\boldsymbol{\beta}}^2}^{(m-1)}) = &-\frac{E_{{\sigma_{\boldsymbol{\beta}}^2}^{(m-1)}}[|\boldsymbol{\beta}|_0]}{2} \log (\sigma_{\boldsymbol{\beta}}^2) \\
	&- \frac{E_{{\sigma_{\boldsymbol{\beta}}^2}^{(m-1)}}\left[\sum_{S:\beta_S \neq 0} (\beta_S - \mu_{\beta_S})^2 / \sigma^2 \right]}{2\sigma_{\boldsymbol{\beta}}^2}
\end{align*}

We then take the derivative of this expression with respect to $\sigma_{\boldsymbol{\beta}}^2$ to find that the maximum occurs at 

\begin{align}
	{\sigma_{\boldsymbol{\beta}}^2}^{(m)} = \frac{E_{{\sigma_{\boldsymbol{\beta}}^2}^{(m-1)}}\left[\sum_{S:\beta_S \neq 0} (\beta_S - \mu_{\beta_S})^2 / \sigma^2\right]}{E_{{\sigma_{\boldsymbol{\beta}}^2}^{(m-1)}}[|\boldsymbol{\beta}|_0]}
\end{align}

Intuitively, to obtain the empirical Bayes estimate, one must run an MCMC chain and update $\sigma_{\boldsymbol{\beta}}^2$ every $T$ MCMC iterations. This process is continued until the estimate of $\sigma_{\boldsymbol{\beta}}^2$ converges, and then the MCMC can be run conditional on this estimated value.

\section{Proofs of lemma 3.1 and theorem 3.2}

Here we investigate the prior probability that a particular interaction term is equal to zero. Our goal is to penalize interactions of higher order more strongly, and the probability that we force them to zero is one component of this strategy. Since the prior probabilities of inclusion into function $k'$ for each covariate are independent, we will investigate, without loss of generality, the probability that $\zeta_{1k'} = \zeta_{2k'} = \dots = \zeta_{jk'} = 1$ and $\zeta_{j+1k'} = \dots = \zeta_{pk'} = 0$. The probability would remain constant if we changed which of the $p$ exposures we chose to be included or excluded from the interaction term, so we will look at the case where the first $j$ terms are included for simplicity. 

\begin{align*}
	& p(\zeta_{1k'} = \zeta_{2k'} = \dots = \zeta_{jk'} = 1, \zeta_{j+1k'} = \dots = \zeta_{pk'} = 0  \vert \tau_1,\dots, \tau_k) = \\
    & \prod_{h=1}^k \big( p(\text{any}(\zeta_{1h},\dots, \zeta_{jh})  = 0) + p(\text{all}(\zeta_{1h},\dots, \zeta_{jh})  = 1 \\
    & \ \ \ \ \ \ \ \ \ \ \text{and any }(\zeta_{(j+1)h},\dots, \zeta_{ph})  = 1)\big) \\
	&= \prod_{h=1}^k \left( (1 - \tau_h^j) +  \tau_h^j (1 - (1 - \tau_h)^{(p-j)}) \right) \\
	&= \prod_{h=1}^k \left( 1 - \tau_h^j(1 - \tau_h)^{(p-j)} \right) \\
    &\text{Then take the expectation over } \tau_h \text{ to get the unconditional probability} \\
	& E \left[ \prod_{h=1}^k \left[ 1 - \tau_h^j(1 - \tau_h)^{(p-j)} \right] \right] \\
	&= \prod_{h=1}^k \left( E_{\tau_h} \left[ 1 - \tau_h^j(1 - \tau_h)^{(p-j)} \right] \right) \\
	&= \prod_{h=1}^k \left( 1 - \int_0^1 \tau_h^{j + M - 1}(1 - \tau_h)^{(p+\gamma-j-1)} d\tau_h \right) \\
	&= \prod_{h=1}^k \left( 1 - \frac{\Gamma(j + M)\Gamma(p + \gamma - j)}{\Gamma(p + \gamma + M)}\right) \\
	&= \left( 1 - \frac{\Gamma(j + M)\Gamma(p + \gamma - j)}{\Gamma(p + \gamma + M)}\right)^k
\end{align*}

\subsection{Probability increases as $j$ increases if $\gamma \geq p$}

It is of interest  how the variable inclusion proabilities vary as a function of $j$, the order of each interaction. Ideally we would want to show that the probability increases as a function of $j$. We don't expect this to happen in general, but we can find conditions at which it does hold. To simplify expressions with the gamma function we restrict $\gamma$ to be an integer, though we expect the results to extend to all real values of $\gamma$. \\

The expression increasing as a function of $j$ is equivalent to $\Gamma(j + M)\Gamma(p + \gamma - j)$ decreasing with $j$, so we show that $\Gamma(j + M)\Gamma(p + \gamma - j)$ is a decreasing function of $j$ under certain conditions. \\

Let $f(j) = \Gamma(j + M)\Gamma(p + \gamma - j) = (j + M - 1)!(p + \gamma - j -1)!$ and let us examine $f(j)/f(j+1)$.

\begin{align*}
	\frac{f(j)}{f(j+1)} &= \frac{(j + M - 1)! (p + \gamma - j -1)!}{(j+M)! (p + \gamma - j -2)!} \\
	&= \frac{p + \gamma - j -1}{j+M}
\end{align*}

We want this ratio to be greater than 1 for all $j \in \{1,\dots,p-1\}$ for the function to be increasing. Clearly this ratio is getting smaller as $j$ gets larger so we can look at the largest value of $j$, which is $p-1$. 

\begin{align*}
	\frac{f(p-1)}{f(p)} &=  \frac{\gamma}{p + M - 1}
\end{align*}

So clearly if we set $\gamma \geq p + M - 1$, we have the desired result. 

\section{Reversibility of updates for variable inclusion matrix}

Here we will show a condition under which our proposed approach to sampling will maintain the desired target distribution. Let $\pi()$ be the conditional density of $\boldsymbol{\zeta}$ given the other model parameters and the data. Next let $K(\boldsymbol{j} \vert \boldsymbol{i})$ be the transition probability of going to $\boldsymbol{\zeta} = \boldsymbol{j}$ from $\boldsymbol{\zeta} = \boldsymbol{i}.$ Our goal is to show that reversibility holds, i.e. that 

$$\pi(\boldsymbol{i}) K(\boldsymbol{j} \vert \boldsymbol{i}) = \pi(\boldsymbol{j}) K(\boldsymbol{i} \vert \boldsymbol{j})$$

Our strategy for updating $\boldsymbol{\zeta}$ is to randomly draw one of the $pk$ elements in the matrix. if setting this element equal to 1 leads to an interaction of any order, then we must also evaluate whether the data could favor this move simply due to there being a lower order interaction. If we are evaluating a two-way interaction, then we must also evaluate whether there are simply main effects of each of the two exposures, otherwise we could incorrectly accept a move to an interaction model and bias our MCMC towards higher-order models. If including this element leads to a three way interaction, then we must evaluate whether in truth there is only the original two-way interaction, and then some additional terms (whether they be main effects or two-way interactions) that include the element we have chosen to update.  This strategy amounts to creating a list of models to evaluate, and then the probability of moving to any one of them is proportional to their conditional posterior probability, $\pi().$ With this in mind, let us first look at $K(\boldsymbol{j} \vert \boldsymbol{i})$:

\begin{align*}
	K(\boldsymbol{j} \vert \boldsymbol{i}) &= P(\boldsymbol{\zeta}_{n+1} = \boldsymbol{j} | \boldsymbol{\zeta}_n = \boldsymbol{i}) \\
	&= \frac{1}{pk} \sum_{m \in M_{ij}} \frac{\pi(\boldsymbol{j})}{\sum_{\boldsymbol{z} \in \zeta^{(m,i)}} \pi(\boldsymbol{z})} \\
	&=  \frac{\pi(\boldsymbol{j})}{pk} \sum_{m \in M_{ij}} \frac{1}{\sum_{\boldsymbol{z} \in \zeta^{(m,i)}} \pi(\boldsymbol{z})} 
\end{align*}

where $M_{ij}$ is the set of all possible updates among the $pk$ possibilities that contain a possible move to state $\boldsymbol{j}$ from state $\boldsymbol{i}$. Further, $\zeta^{(m,i)}$ denotes the set of all $\boldsymbol{\zeta}$ matrices that are considered if we update $m$ and we are currently at state $\boldsymbol{i}$. Now let's put this back into the equation for reversibility:

\begin{align*}
	& \pi(\boldsymbol{i}) K(\boldsymbol{j} \vert \boldsymbol{i}) = \pi(\boldsymbol{j}) K(\boldsymbol{i} \vert \boldsymbol{j}) \\
	\leftrightarrow &\pi(\boldsymbol{i}) \frac{\pi(\boldsymbol{j})}{pk} \sum_{m \in M_{ij}} \frac{1}{\sum_{\boldsymbol{z} \in \zeta^{(m,i)}} \pi(\boldsymbol{z})}  = \pi(\boldsymbol{j}) \frac{\pi(\boldsymbol{i})}{pk} \sum_{m \in M_{ji}} \frac{1}{\sum_{\boldsymbol{z} \in \zeta^{(m,j)}} \pi(\boldsymbol{z})}  \\
	\leftrightarrow & \sum_{m \in M_{ij}} \frac{1}{\sum_{\boldsymbol{z} \in \zeta^{(m,i)}} \pi(\boldsymbol{z})}  = \sum_{m \in M_{ji}} \frac{1}{\sum_{\boldsymbol{z} \in \zeta^{(m,j)}} \pi(\boldsymbol{z})}.
\end{align*}

This equality must hold if two conditions hold. First $M_{ij} = M_{ji}$, i.e. the updates that can take you from state $\boldsymbol{i}$ to state $\boldsymbol{j}$ are the same updates that can take you in the other direction. Second, we need $\zeta^{(m,i)} = \zeta^{(m,j)}$, i.e. the space of models that are being considered if we update element $m$ and are currently at state $\boldsymbol{i}$ is the same as those if we are at state $\boldsymbol{j}$ and we update element $m$. Thankfully we are in full control of the models that we choose to evaluate if we choose to update element $m$, so we can ensure this condition holds. Our first priority is to ensure that we consider all models that contain lower-order interactions as described in the manuscript so that we don't incorrectly move to a higher-order interaction when the truth is a lower-order model. Next, we will do a check to ensure that the required conditions hold, and if not, we will add any other models to evaluate that will make the conditions satisfied. If the model is complex and includes a large number of interactions, then this search can be computationally intensive, so we detail a strategy below using Metropolis Hastings that does not suffer from the same computational challenges, yet also satisfies reversibility.

\subsection{Metropolis Hastings updates for variable inclusion matrix}

Here we describe a strategy for proposing models and moving around the model space using Metropolis Hastings. As above, we will again choose one of the $pk$ elements of the matrix randomly. If setting this element equal to 1 leads to an interaction of any order, then we will create a list of possible models that could lead to us choosing this model that includes all lower-order interactions. We then choose one of these models randomly and use it as a proposed new variable inclusion matrix. For a Metropolis Hastings step, we simply need the density $\pi()$ and the proposal probabilities, $Q(\boldsymbol{j} \vert \boldsymbol{i})$ of going from state $\boldsymbol{i}$ to $\boldsymbol{j}$. The proposal probability is straightforward to calculate here as it is simply one divided by the number of models in the list. These proposal probabilities can easily be calculated in the reverse direction as well to find the probability of going from the proposed model to the current model, and then the standard acceptance probability $$\text{min} \bigg \{ 1, \frac{\pi(\boldsymbol{j}) Q(\boldsymbol{i} \vert \boldsymbol{j})}{\pi(\boldsymbol{i}) Q(\boldsymbol{j} \vert \boldsymbol{i})} \bigg\}$$

\noindent can be used. By being a Metropolis Hastings step this is guaranteed to satisfy reversibility, and it should not get stuck in modes with unnecessary higher-order interactions because it considers all models that could lead to the data favoring a higher-order move. 

\section{MCMC convergence in the Bangladesh data}

\begin{figure}[H]
\centering
	  \includegraphics[width=0.4\linewidth]{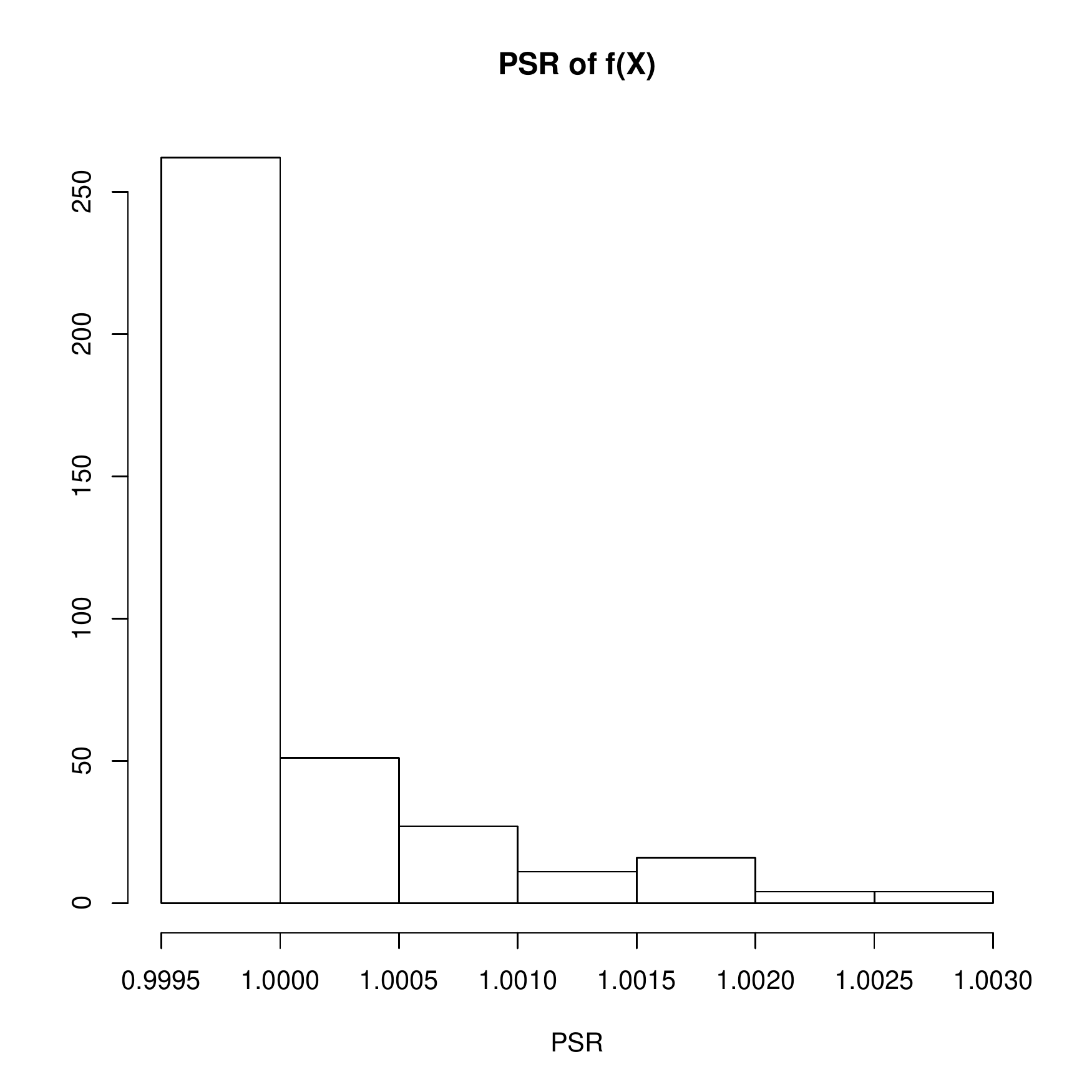}
\caption{Histogram of PSR values for $f(X_i)$ in the Bangladesh data.}
\label{fig:PSRbangladesh}
\end{figure}

\begin{figure}[H]
\centering
	  \includegraphics[width=0.65\linewidth]{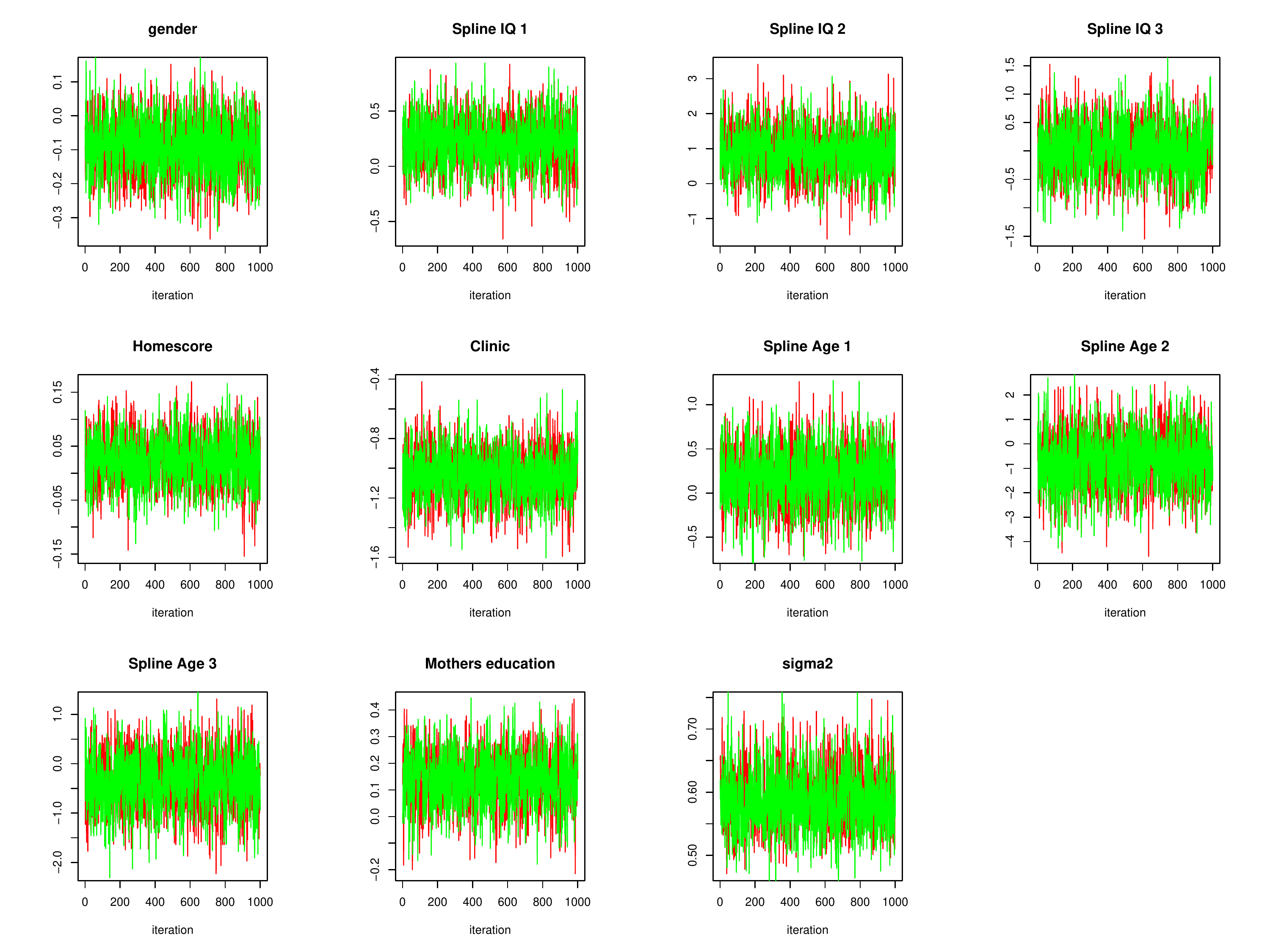}
\caption{Trace plots for $\boldsymbol{\beta}_C$ and $\sigma^2$ in the Bangladesh data. }
\label{fig:TRACEbangladesh}
\end{figure}

To assess convergence in the Bangladesh data set we kept track of the potential scale reduction factor (PSR, \cite{gelman2014bayesian}) for all identifiable parameters in our model. Therefore we examined $f(X_i)$ for $i = 1, \dots, n$, all elements of $\boldsymbol{\beta}_C$, and $\sigma^2$. We found that the PSR value was below 1.005 for all parameters in our model, which leads us to believe that our MCMC has converged. A plot of the PSR values for $f(X_i)$ can be found in Figure \ref{fig:PSRbangladesh}. Subsequently, we show the trace plots for all parameters in $\boldsymbol{\beta}_C$ and $\sigma^2$ in Figure \ref{fig:TRACEbangladesh} and find that our chain appears to be converging. 

\section{MCMC convergence in the NHANES data}

\begin{figure}[H]
\centering
	  \includegraphics[width=0.4\linewidth]{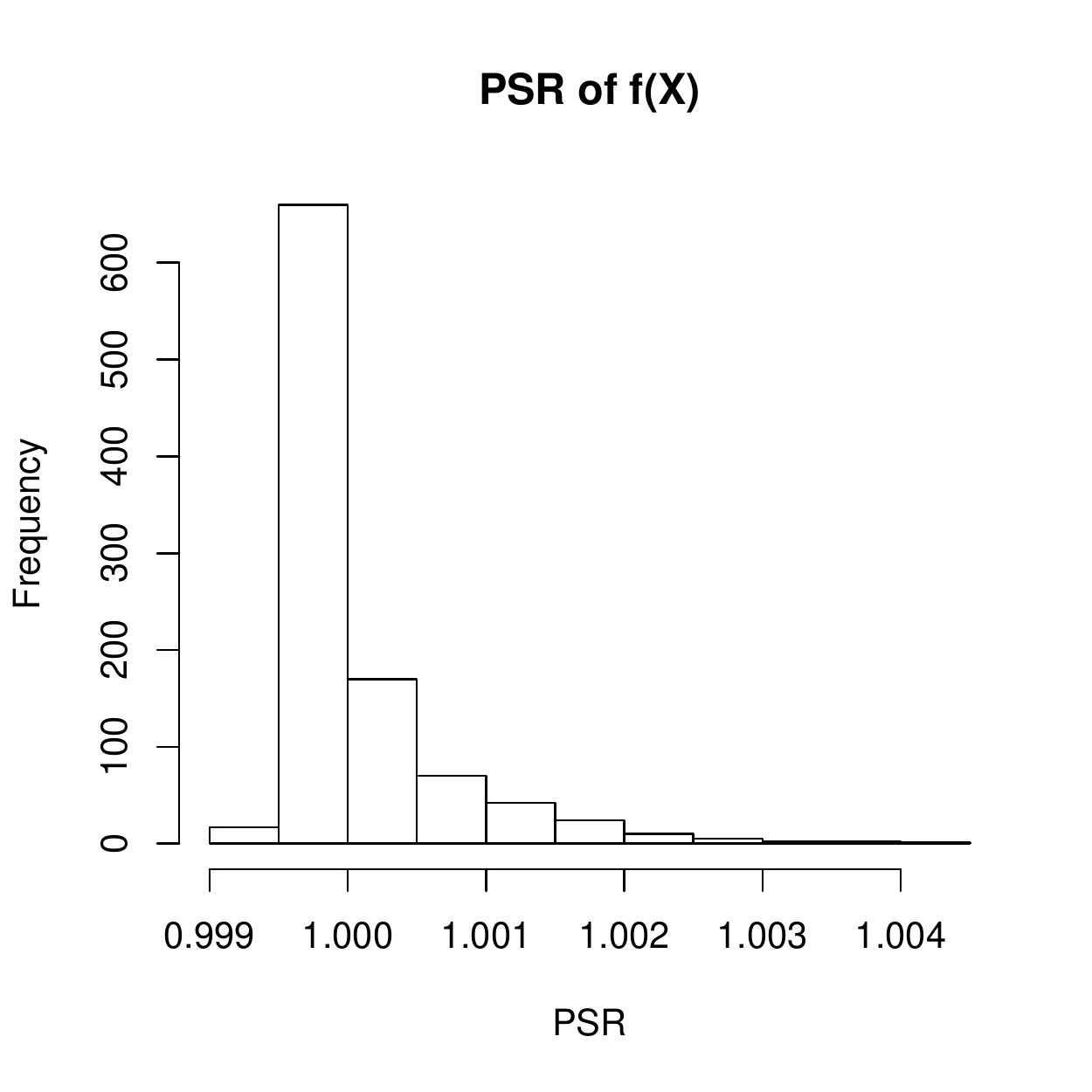}
\caption{Histogram of PSR values for $f(X_i)$ in the NHANES data.}
\label{fig:PSRnhanes}
\end{figure}

\begin{figure}[H]
\centering
	  \includegraphics[width=0.65\linewidth]{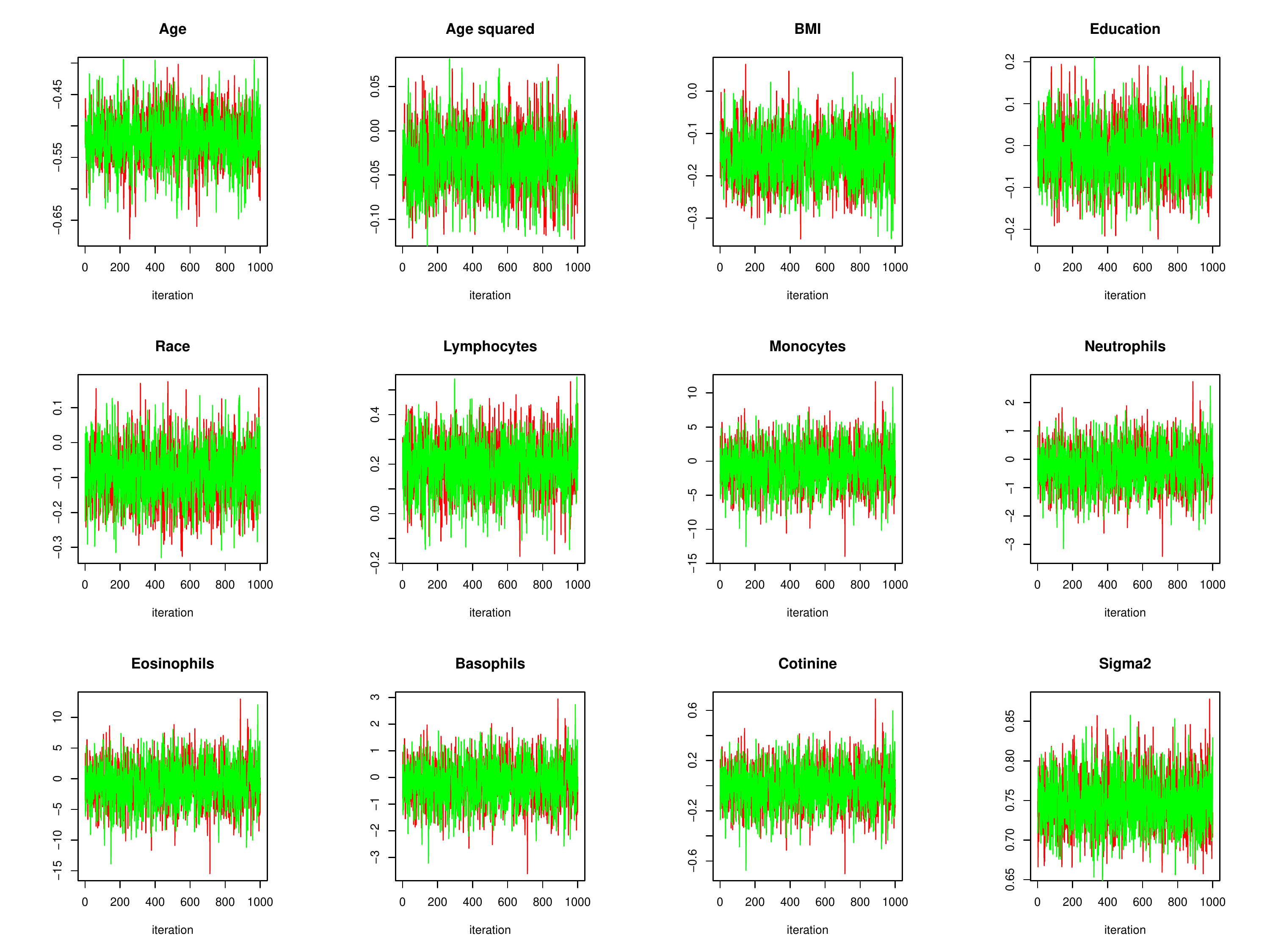}
\caption{Trace plots for $\boldsymbol{\beta}_C$ and $\sigma^2$ in the NHANES data. }
\label{fig:TRACEnhanes}
\end{figure}

To assess convergence in the NHANES data set we kept track of the PSR for all identifiable parameters in our model. Therefore we examined $f(X_i)$ for $i = 1, \dots, n$, all elements of $\boldsymbol{\beta}_C$, and $\sigma^2$. We found that the PSR value was below 1.005 for all parameters in our model, which leads us to believe that our MCMC has converged. A plot of the PSR values for $f(X_i)$ can be found in Figure \ref{fig:PSRnhanes}. Subsequently, we show the trace plots for a subset of parameters in $\boldsymbol{\beta}_C$ and $\sigma^2$ in Figure \ref{fig:TRACEnhanes} and find that our chain appears to be converging. 

\bibliographystyle{authordate1}
\bibliography{MixtureFactorization}

\begin{thebibliography}{}

\bibitem[\protect\citename{Bayley, }2006]{bayley2006bayley}
Bayley, Nancy. 2006.
\newblock {\em Bayley Scales of Infant and Toddler Development-Third Edition:
  Administration manual}.

\bibitem[\protect\citename{Bengio {\em et~al.\ }\relax,
  }2003]{bengio2003neural}
Bengio, Yoshua, Ducharme, R{\'e}jean, Vincent, Pascal, \& Jauvin, Christian.
  2003.
\newblock A neural probabilistic language model.
\newblock {\em Journal of machine learning research}, {\bf 3}(Feb), 1137--1155.

\bibitem[\protect\citename{Bien {\em et~al.\ }\relax, }2013]{bien2013lasso}
Bien, Jacob, Taylor, Jonathan, \& Tibshirani, Robert. 2013.
\newblock A lasso for hierarchical interactions.
\newblock {\em Annals of statistics}, {\bf 41}(3), 1111.

\bibitem[\protect\citename{Bobb {\em et~al.\ }\relax, }2014]{Bobb2014}
Bobb, Jennifer~F., Valeri, Linda, {Claus Henn}, Birgit, Christiani, David~C.,
  Wright, Robert~O., Mazumdar, Maitreyi, Godleski, John~J., \& Coull, Brent~A.
  2014.
\newblock {Bayesian kernel machine regression for estimating the health effects
  of multi-pollutant mixtures}.
\newblock {\em Biostatistics}, {\bf 16}(3), 493--508.

\bibitem[\protect\citename{Bobb {\em et~al.\ }\relax,
  }2018]{bobb2018statistical}
Bobb, Jennifer~F, Henn, Birgit~Claus, Valeri, Linda, \& Coull, Brent~A. 2018.
\newblock Statistical software for analyzing the health effects of multiple
  concurrent exposures via Bayesian kernel machine regression.
\newblock {\em Environmental Health}, {\bf 17}(1), 67.

\bibitem[\protect\citename{Braun, }2017]{braun2017early}
Braun, Joseph~M. 2017.
\newblock Early-life exposure to EDCs: role in childhood obesity and
  neurodevelopment.
\newblock {\em Nature Reviews Endocrinology}, {\bf 13}(3), 161.

\bibitem[\protect\citename{Braun {\em et~al.\ }\relax, }2016]{Braun2016}
Braun, Joseph~M., Gennings, Chris, Hauser, Russ, \& Webster, Thomas~F. 2016.
\newblock {What can epidemiological studies tell us about the impact of
  chemical mixtures on human health?}
\newblock {\em Environmental Health Perspectives}, {\bf 124}(1), A6--A9.

\bibitem[\protect\citename{Breiman, }2001]{breiman2001random}
Breiman, Leo. 2001.
\newblock Random forests.
\newblock {\em Machine learning}, {\bf 45}(1), 5--32.

\bibitem[\protect\citename{Carlin {\em et~al.\ }\relax,
  }2013]{carlin2013unraveling}
Carlin, Danielle~J, Rider, Cynthia~V, Woychik, Rick, \& Birnbaum, Linda~S.
  2013.
\newblock Unraveling the health effects of environmental mixtures: an NIEHS
  priority.
\newblock {\em Environmental health perspectives}, {\bf 121}(1), a6.

\bibitem[\protect\citename{Casella, }2001]{casella2001empirical}
Casella, George. 2001.
\newblock Empirical bayes gibbs sampling.
\newblock {\em Biostatistics}, {\bf 2}(4), 485--500.

\bibitem[\protect\citename{Chipman {\em et~al.\ }\relax,
  }2010]{chipman2010bart}
Chipman, Hugh~A, George, Edward~I, McCulloch, Robert~E, {\em et~al.\ }\relax.
  2010.
\newblock BART: Bayesian additive regression trees.
\newblock {\em The Annals of Applied Statistics}, {\bf 4}(1), 266--298.

\bibitem[\protect\citename{Cristianini \& Shawe-Taylor,
  }2000]{cristianini2000introduction}
Cristianini, Nello, \& Shawe-Taylor, John. 2000.
\newblock {\em An introduction to support vector machines and other
  kernel-based learning methods}.
\newblock Cambridge university press.

\bibitem[\protect\citename{Gelman {\em et~al.\ }\relax,
  }2008]{gelman2008weakly}
Gelman, Andrew, Jakulin, Aleks, Pittau, Maria~Grazia, \& Su, Yu-Sung. 2008.
\newblock A weakly informative default prior distribution for logistic and
  other regression models.
\newblock {\em The Annals of Applied Statistics},  1360--1383.

\bibitem[\protect\citename{Gelman {\em et~al.\ }\relax,
  }2014]{gelman2014bayesian}
Gelman, Andrew, Carlin, John~B, Stern, Hal~S, \& Rubin, Donald~B. 2014.
\newblock {\em Bayesian data analysis}.
\newblock  Vol. 2.
\newblock Chapman \& Hall/CRC Boca Raton, FL, USA.

\bibitem[\protect\citename{George \& McCulloch, }1993]{george1993variable}
George, Edward~I, \& McCulloch, Robert~E. 1993.
\newblock Variable selection via Gibbs sampling.
\newblock {\em Journal of the American Statistical Association}, {\bf 88}(423),
  881--889.

\bibitem[\protect\citename{Guo {\em et~al.\ }\relax, }2016]{guo2016boosting}
Guo, Fangjian, Wang, Xiangyu, Fan, Kai, Broderick, Tamara, \& Dunson, David~B.
  2016.
\newblock Boosting Variational Inference.
\newblock {\em arXiv preprint arXiv:1611.05559}.

\bibitem[\protect\citename{Hahn {\em et~al.\ }\relax, }2017]{hahn2017bayesian}
Hahn, P~Richard, Murray, Jared, \& Carvalho, Carlos~M. 2017.
\newblock Bayesian regression tree models for causal inference: regularization,
  confounding, and heterogeneous effects.
\newblock {\em Confounding, and Heterogeneous Effects (October 5, 2017)}.

\bibitem[\protect\citename{Hahn {\em et~al.\ }\relax,
  }2018]{hahn2018regularization}
Hahn, P~Richard, Carvalho, Carlos~M, Puelz, David, He, Jingyu, {\em et~al.\
  }\relax. 2018.
\newblock Regularization and confounding in linear regression for treatment
  effect estimation.
\newblock {\em Bayesian Analysis}, {\bf 13}(1), 163--182.

\bibitem[\protect\citename{Hao {\em et~al.\ }\relax, }2014]{Hao2014}
Hao, Ning, Feng, Yang, \& Zhang, Hao~Helen. 2014.
\newblock {Model Selection for High Dimensional Quadratic Regression via
  Regularization}.
\newblock {\bf 85721}, 26.

\bibitem[\protect\citename{Harley {\em et~al.\ }\relax,
  }2017]{harley2017association}
Harley, Kim~G, Berger, Kimberly, Rauch, Stephen, Kogut, Katherine, Henn,
  Birgit~Claus, Calafat, Antonia~M, Huen, Karen, Eskenazi, Brenda, \& Holland,
  Nina. 2017.
\newblock Association of prenatal urinary phthalate metabolite concentrations
  and childhood BMI and obesity.
\newblock {\em Pediatric research}, {\bf 82}(3), 405.

\bibitem[\protect\citename{Henn {\em et~al.\ }\relax, }2010]{henn2010early}
Henn, Birgit~Claus, Ettinger, Adrienne~S, Schwartz, Joel, T{\'e}llez-Rojo,
  Martha~Mar{\'\i}a, Lamadrid-Figueroa, H{\'e}ctor, Hern{\'a}ndez-Avila,
  Mauricio, Schnaas, Lourdes, Amarasiriwardena, Chitra, Bellinger, David~C, Hu,
  Howard, {\em et~al.\ }\relax. 2010.
\newblock Early postnatal blood manganese levels and children’s
  neurodevelopment.
\newblock {\em Epidemiology (Cambridge, Mass.)}, {\bf 21}(4), 433.

\bibitem[\protect\citename{Henn {\em et~al.\ }\relax, }2014]{henn2014chemical}
Henn, Birgit~Claus, Coull, Brent~A, \& Wright, Robert~O. 2014.
\newblock Chemical mixtures and children’s health.
\newblock {\em Current opinion in pediatrics}, {\bf 26}(2), 223.

\bibitem[\protect\citename{Kortenkamp {\em et~al.\ }\relax,
  }2007]{kortenkamp2007low}
Kortenkamp, Andreas, Faust, Michael, Scholze, Martin, \& Backhaus, Thomas.
  2007.
\newblock Low-level exposure to multiple chemicals: reason for human health
  concerns?
\newblock {\em Environmental Health Perspectives}, {\bf 115}(Suppl 1),
  106--114.

\bibitem[\protect\citename{Lazarevic {\em et~al.\ }\relax,
  }2019]{lazarevic2019statistical}
Lazarevic, Nina, Barnett, Adrian~G, Sly, Peter~D, \& Knibbs, Luke~D. 2019.
\newblock Statistical methodology in studies of prenatal exposure to mixtures
  of endocrine-disrupting chemicals: a review of existing approaches and new
  alternatives.
\newblock {\em Environmental health perspectives}, {\bf 127}(2), 026001.

\bibitem[\protect\citename{Lim \& Hastie, }2015]{lim2015learning}
Lim, Michael, \& Hastie, Trevor. 2015.
\newblock Learning interactions via hierarchical group-lasso regularization.
\newblock {\em Journal of Computational and Graphical Statistics}, {\bf 24}(3),
  627--654.

\bibitem[\protect\citename{Miller {\em et~al.\ }\relax,
  }2016]{miller2016variational}
Miller, Andrew~C, Foti, Nicholas, \& Adams, Ryan~P. 2016.
\newblock Variational Boosting: Iteratively Refining Posterior Approximations.
\newblock {\em arXiv preprint arXiv:1611.06585}.

\bibitem[\protect\citename{Mitchell \& Beauchamp, }1988]{mitchell1988bayesian}
Mitchell, Toby~J, \& Beauchamp, John~J. 1988.
\newblock Bayesian variable selection in linear regression.
\newblock {\em Journal of the American Statistical Association}, {\bf 83}(404),
  1023--1032.

\bibitem[\protect\citename{Mitro {\em et~al.\ }\relax, }2015]{mitro2015cross}
Mitro, Susanna~D, Birnbaum, Linda~S, Needham, Belinda~L, \& Zota, Ami~R. 2015.
\newblock Cross-sectional associations between exposure to persistent organic
  pollutants and leukocyte telomere length among US adults in NHANES,
  2001--2002.
\newblock {\em Environmental health perspectives}, {\bf 124}(5), 651--658.

\bibitem[\protect\citename{O'Hagan \& Kingman, }1978]{o1978curve}
O'Hagan, Anthony, \& Kingman, JFC. 1978.
\newblock Curve fitting and optimal design for prediction.
\newblock {\em Journal of the Royal Statistical Society. Series B
  (Methodological)},  1--42.

\bibitem[\protect\citename{Qamar \& Tokdar, }2014]{Qamar2014}
Qamar, Shaan, \& Tokdar, ST. 2014.
\newblock {Additive Gaussian Process Regression}.
\newblock {\em arXiv preprint arXiv:1411.7009}, ~28.

\bibitem[\protect\citename{Radchenko \& James, }2010]{radchenko2010variable}
Radchenko, Peter, \& James, Gareth~M. 2010.
\newblock Variable selection using adaptive nonlinear interaction structures in
  high dimensions.
\newblock {\em Journal of the American Statistical Association}, {\bf
  105}(492), 1541--1553.

\bibitem[\protect\citename{Reich {\em et~al.\ }\relax,
  }2009]{reich2009variable}
Reich, Brian~J, Storlie, Curtis~B, \& Bondell, Howard~D. 2009.
\newblock Variable selection in Bayesian smoothing spline ANOVA models:
  Application to deterministic computer codes.
\newblock {\em Technometrics}, {\bf 51}(2), 110--120.

\bibitem[\protect\citename{Ro{\v{c}}kov{\'a} \& George,
  }2016]{rovckova2016spike}
Ro{\v{c}}kov{\'a}, Veronika, \& George, Edward~I. 2016.
\newblock The spike-and-slab lasso.
\newblock {\em Journal of the American Statistical Association}.

\bibitem[\protect\citename{Scheipl {\em et~al.\ }\relax,
  }2012]{scheipl2012spike}
Scheipl, Fabian, Fahrmeir, Ludwig, \& Kneib, Thomas. 2012.
\newblock Spike-and-slab priors for function selection in structured additive
  regression models.
\newblock {\em Journal of the American Statistical Association}, {\bf
  107}(500), 1518--1532.

\bibitem[\protect\citename{Shively {\em et~al.\ }\relax,
  }1999]{shively1999variable}
Shively, Thomas~S, Kohn, Robert, \& Wood, Sally. 1999.
\newblock Variable selection and function estimation in additive nonparametric
  regression using a data-based prior.
\newblock {\em Journal of the American Statistical Association}, {\bf 94}(447),
  777--794.

\bibitem[\protect\citename{Taylor {\em et~al.\ }\relax,
  }2016]{taylor2016statistical}
Taylor, Kyla~W, Joubert, Bonnie~R, Braun, Joe~M, Dilworth, Caroline, Gennings,
  Chris, Hauser, Russ, Heindel, Jerry~J, Rider, Cynthia~V, Webster, Thomas~F,
  \& Carlin, Danielle~J. 2016.
\newblock Statistical approaches for assessing health effects of environmental
  chemical mixtures in epidemiology: lessons from an innovative workshop.
\newblock {\em Environmental health perspectives}, {\bf 124}(12), A227.

\bibitem[\protect\citename{Tibshirani, }1996]{tibshirani1996regression}
Tibshirani, Robert. 1996.
\newblock Regression shrinkage and selection via the lasso.
\newblock {\em Journal of the Royal Statistical Society. Series B
  (Methodological)},  267--288.

\bibitem[\protect\citename{Valeri {\em et~al.\ }\relax, }2017]{valeri2017joint}
Valeri, Linda, Mazumdar, Maitreyi~M, Bobb, Jennifer~F, Henn, Birgit~Claus,
  Rodrigues, Ema, Sharif, Omar~IA, Kile, Molly~L, Quamruzzaman, Quazi, Afroz,
  Sakila, Golam, Mostafa, {\em et~al.\ }\relax. 2017.
\newblock The joint effect of prenatal exposure to metal mixtures on
  neurodevelopmental outcomes at 20--40 months of age: Evidence from rural
  Bangladesh.
\newblock {\em Environmental health perspectives}, {\bf 125}(6).

\bibitem[\protect\citename{Watanabe, }2010]{watanabe2010asymptotic}
Watanabe, Sumio. 2010.
\newblock Asymptotic equivalence of Bayes cross validation and widely
  applicable information criterion in singular learning theory.
\newblock {\em Journal of Machine Learning Research}, {\bf 11}(Dec),
  3571--3594.

\bibitem[\protect\citename{Wood {\em et~al.\ }\relax, }2002]{wood2002model}
Wood, Sally, Kohn, Robert, Shively, Tom, \& Jiang, Wenxin. 2002.
\newblock Model selection in spline nonparametric regression.
\newblock {\em Journal of the Royal Statistical Society: Series B (Statistical
  Methodology)}, {\bf 64}(1), 119--139.

\bibitem[\protect\citename{Wood, }2006]{wood2006low}
Wood, Simon~N. 2006.
\newblock Low-Rank Scale-Invariant Tensor Product Smooths for Generalized
  Additive Mixed Models.
\newblock {\em Biometrics}, {\bf 62}(4), 1025--1036.

\bibitem[\protect\citename{Yau {\em et~al.\ }\relax, }2003]{yau2003bayesian}
Yau, Paul, Kohn, Robert, \& Wood, Sally. 2003.
\newblock Bayesian variable selection and model averaging in high-dimensional
  multinomial nonparametric regression.
\newblock {\em Journal of Computational and Graphical Statistics}, {\bf 12}(1),
  23--54.

\bibitem[\protect\citename{Zhao {\em et~al.\ }\relax, }2009]{zhao2009composite}
Zhao, Peng, Rocha, Guilherme, \& Yu, Bin. 2009.
\newblock The composite absolute penalties family for grouped and hierarchical
  variable selection.
\newblock {\em The Annals of Statistics},  3468--3497.

\bibitem[\protect\citename{Zhou {\em et~al.\ }\relax, }2014]{Zhou2014}
Zhou, Jing, Bhattacharya, Anirban, Herring, Amy~H., \& Dunson, David~B. 2014.
\newblock {Bayesian factorizations of big sparse tensors}.
\newblock {\em Journal of the American Statistical Association}, {\bf
  1459}(1994), 00--00.

\bibitem[\protect\citename{Zou, }2006]{zou2006adaptive}
Zou, Hui. 2006.
\newblock The adaptive lasso and its oracle properties.
\newblock {\em Journal of the American statistical association}, {\bf
  101}(476), 1418--1429.

\bibitem[\protect\citename{Zou \& Hastie, }2005]{zou2005regularization}
Zou, Hui, \& Hastie, Trevor. 2005.
\newblock Regularization and variable selection via the elastic net.
\newblock {\em Journal of the Royal Statistical Society: Series B (Statistical
  Methodology)}, {\bf 67}(2), 301--320.

\end{thebibliography}

\end{document}